# An Exploratory Study of Multimodal Physiological Data in Jazz Improvisation Using Basic Machine Learning Techniques

By

Yawen Zhang

2023

A Dissertation Presented to The University of Nottingham

In part Consideration for the Degree of "MSc Computer Science"

# Abstract

Music performance, akin to human society, epitomizes a multifaceted realm characterized by intricate interplays between physiological and psychological responses. Despite undergoing analogous training, no two musicians manifest identical reactions during a performance. Such individual disparities are influenced by a myriad of factors ranging from genetic makeup, past experiences, pedagogical methods to the everyday stresses of life.

Electrodermal Activity (EDA), also known as Galvanic Skin Response (GSR), stands out as a quintessential indicator of the autonomic nervous system's status. As a marker for skin conductance, it reflects variations in skin moisture, thereby offering insights into an individual's physiological condition to some extent. Governed by the sympathetic nervous system, EDA is perceived as an index of physiological or psychological arousal. In our investigation, EDA serves as the predominant physiological measure, shedding light on musicians' physiological responses during improvisational performances.

Notably, genetic variations might account for differences in muscular structure, neural pathways, or even the body's response to stress. A classical musician might adopt a disparate approach to a composition compared to a jazz counterpart, attributed to their training backgrounds. Techniques, postures, and even emotional connections with music can vary, resulting in differential physiological responses. Moreover, a musician's psychological state can profoundly shape their physiological reactions. Individual stress, mood, and even levels of motivation might give rise to distinct physiological patterns.

Our study delves into the "Embodied Musicking Dataset," exploring the intertwined relationships and correlations between physiological and psychological dimensions during improvisational music performances. The primary objective is to ascertain the presence of a definitive causal or correlational relationship between these states and comprehend their manifestation in musical compositions. This rich dataset provides a perspective on how musicians coordinate their physicality with sonic events in real-time improvisational scenarios, emphasizing the concept of "Embodied Musicking."

# Contents

# Introduction

Music, as a carrier of culture and emotion, has an inseparable bond with human existence. From ancient court music to modern electronic music, music not only serves as an emotional outlet but also has a profound connection with our cognition, behavior, and physiological state. This deep relationship between music and human life has attracted the attention of scientists, leading them to delve into how music interacts with our brain, emotions, and physiological conditions.

With the advancement of technology and research methodologies, studies on the relationship between music and human physiological responses have deepened. For instance, some research has found that music can activate specific areas of the brain associated with emotions, memory, and concentration. Additionally, studies have shown that music can influence our heart rate, respiration, and blood pressure, further confirming that music not only affects our psychological state but also has a tangible connection to our physiological state.

The state of flow, as a unique psychological experience, holds particular importance in the realm of music. This state represents complete immersion in an activity, where one loses track of time and is fully engrossed in the task at hand. Music performance, requiring high levels of concentration and skill, offers musicians an ideal setting to experience this flow state. Thus, an in-depth study of musicians' physiological and psychological responses during flow is crucial for understanding music creation.

However, despite extensive research on music's influence on physiology and psychology, there remain limitations. Many studies have solely focused on the music experience, with limited research specifically looking into the intricate relationship between music performance and physiological responses. Moreover, how to apply these research findings in real-world music-related domains remains an unresolved question.

Given this backdrop, this study aims to explore the physiological reactions of musicians during their performance, especially when they are in a flow state. Through a comprehensive analysis of physiological data, musical performance, and the actual experiences of musicians (flow), we hope to provide new perspectives and insights into the complex interplay between music, physiology, and psychology.

# Literature Review

## Correlation Analysis: Physiological Data and Psychological State

### Autonomic Neural Activity and Psychological State

There is believed to be a causal relationship between physiological and psychological states. When individuals face a stimulus, they manifest an internal emotional state, which is elicited by the activation of a series of neurons in the brain, termed emotions. Experiencing such an emotional state often leads to physiological changes, such as increased heart rate, sweating, and facial expressions. These changes are triggered by the activation of neurons in the brain, implying an unconscious neural response(Bosse et al., 2008). Autonomic neural activity corresponds to physiological expressions under the control of the ANS. The term "autonomic" suggests that this part of the nervous system functions independently of any voluntary neural or cognitive control (Sequeira et al., 2009) and is a response to regulate internal and external stimuli. The autonomic nervous system plays a pivotal role in emotional responses, including cardiovascular, respiratory, and galvanic skin responses, governing numerous physiological processes such as heart rate, blood pressure, respiration, and sweating. These physiological responses are often employed as indicators of emotional arousal and can be assessed using various physiological measuring methods(Kreibig, 2010). By measuring autonomic neural activity, we can obtain a relatively comprehensive understanding of brain functions, not just focusing on the activity of specific areas. This approach aids in understanding the brain's response in various situations, such as dealing with stress, emotions, and cognitive tasks. It's essential to note, however, that the measurement of autonomic neural activity is indirect as they only provide some indicators of brain activity, not a direct measurement. Consequently, the interpretation of this method requires a combination of other methods and theories (Venables, 1991).

### Importance and Diversity of Electrodermal Activity (EDA)

Electrodermal Activity (EDA), also known as Galvanic Skin Response (GSR), is one of the fundamental indicators of the state of the autonomic nervous system(Sarchiapone et al., 2018). Serving as a measure of skin conductivity, it reflects changes in skin moisture, thereby revealing an individual's physiological state to some extent(Bailey, 2017). As EDA is regulated by the sympathetic nervous system, it is considered an indicator of physiological or psychological arousal(Shukla et al., 2021). In this study, we employ EDA as a physiological measurement method to investigate the physiological responses of musicians during improvisational performances.

While EDA has been extensively utilized in psychological research to assess everything from basic emotional reactions to intricate cognitive processes, its application in the domain of music psychology is relatively limited. Given that music is a potent emotional stimulus, this research gap is particularly conspicuous. In improvisational musical performances, musicians are tasked with processing a vast amount of emotional and cognitive information, which might be reflected in the EDA readings. Consequently, this study aims to explore the application of EDA in the realm of music and its pivotal role in the performances of musicians.

Building upon previous research, we aspire to conduct a comprehensive analysis of the physiological data and the performance of musicians, delving deep into the intricate relationship between music performance, physiology, and psychology, offering fresh insights and research directions for music education, music therapy, and other related fields.

## EEG and Its Application in Musicians

Electroencephalography (EEG) offers a non-invasive method to measure changes in the electrical potentials of brain activity(Lin et al., 2010). These changes mirror the activity of cortical neurons in the brain. Depending on the electrode placement and the specific activity of the captured brain region, different EEG channels can provide invaluable information about brain function and structure.

T3 (Left Temporal Lobe): The temporal lobe is the primary area in our brain for processing auditory information and is also involved in memory, emotion, and some linguistic functions. During musical performance, activity in the T3 might increase, especially when musicians deal with complex melodies or harmonies.

T4 (Right Temporal Lobe): Although symmetrically positioned to T3 in the brain, the right temporal lobe might function slightly differently from the left when processing certain musical and linguistic tasks. Some studies suggest that the right temporal lobe might be more active in processing musical melodies and pitches.

O1 & O2 (Occipital Lobe): These channels reflect the activity of the brain's visual processing centers. In musical performances, especially those that involve reading musical scores or interacting with visual stimuli, the activity of these channels might increase.(Harris et al., 2018)

In this study, considering the uniqueness of musicians during improvisational performances, we anticipate observing specific activity related to auditory processing and emotional expression in the T3 and T4 channels. Simultaneously, although O1 and O2 are primarily associated with visual processing, in some improvisational performances, musicians might rely on their visual memory or visual interaction with other instrumentalists, possibly influencing the activity of these two channels.

It's noteworthy that while EEG provides a direct measurement of brain activity, accurate interpretation and understanding of these signals still demand profound expertise. More importantly, for more comprehensive and accurate conclusions, we need to integrate EEG data with other physiological and behavioral data, keeping in mind the specific conditions and design of the experiment.

## Movement Data Analysis

The movements and postures of the human body are reflective of its physiological and psychological states (Honing, 2003). Analyzing movement data, especially the movements and postures of musicians during performances, can offer us profound insights into their skills, emotional, and cognitive states. This analysis typically encompasses multiple facets:

1. Speed and Rhythm of Movement: The variations in the speed and rhythm of movement might correlate with a musician's skill level, emotional state, and cognitive load. For instance, highly skilled musicians might exhibit more stable and synchronized movement rhythms, while their movement speed might increase when confronted with challenging sections or emotionally charged parts.

2. Range and Pattern of Movement: By analyzing the range and pattern of a musician's movements, we can understand how they utilize their body during a performance. Certain movement patterns might be associated with specific emotional expressions or musical styles.

3. Dynamic Stability: Maintaining stability and balance is crucial during a musical performance. By studying the dynamic stability of musicians, we can understand how they adjust and adapt during a performance and how they might psychologically and physiologically interact with the music.

4. Interaction between Body and Instrument: Analyzing how musicians interact with their instruments can offer insights into their technical prowess and performance style. For instance, specific bodily movements might correlate with particular musical techniques or emotional expressions.

In essence, movement data analysis provides us with an intuitive and quantitative method to comprehend the bodily movements and postures of musicians during their performances. Coupled with physiological and psychological measurements, this analysis can present us with a holistic perspective to understand the intricate interplay between music, the body, and the brain. In this study, we will integrate movement data with the previously mentioned physiological measurements to explore the holistic responses of musicians during a state of flow.

# Flow Theory: Peak Experiences in Musical Performance

The concept of 'flow,' first introduced by psychologist Mihaly Csikszentmihalyi (Csikszentmihalyi et al., 2014), describes the state of complete immersion and focus an individual experiences in a particular activity. In this state, external distractions and self-consciousness are momentarily suppressed, allowing the individual to experience deep concentration, satisfaction, and joy.(Snyder and Lopez, 2001)

In the realm of music, the state of flow holds unique significance. Numerous musicians describe their most memorable and outstanding performances as being closely linked to experiencing the state of flow(Wrigley and Emmerson, 2013). Based on this observation, our study aims to delve deeply into whether there are physiological responses related to the state of flow in musicians during their performances.

The state of flow is universally regarded as a high-level experience in musical performances. In this state, not only are musicians technically adept, but they also reach the pinnacle in emotional and creative expression(Diaz, 2013). However, the physiological mechanisms and markers accompanying this remain unclear. Our research aims to bridge this knowledge gap, seeking to identify whether specific physiological indicators correspond to musical performances in a state of flow.

The relationship between flow and high-level performances, although extensively studied in the field of psychology, is relatively under-researched in the domain of music, especially concerning physiological responses related to flow. Delving deep into this area can not only help us understand the intrinsic mechanisms of musical performances more holistically but might also offer novel directions and strategies for music education and practice. For instance, by identifying physiological markers associated with the state of flow, music educators and performers can fine-tune their training and performance strategies, making it easier to attain this ideal state.

# Methodology

## Background and Objectives

This project aims to delve deeply into the Embodied Musicking Dataset, exploring the interaction and correlation between physiological and psychological dimensions during improvisational musical performances. The core objective is to discern if there's a distinct causal or correlational relationship between these two states and attempt to elucidate how these relationships manifest in musical composition.

# Data Selection

## Multidimensionality of the Dataset

The uniqueness of this dataset lies in its multidimensionality, integrating various physiological and psychological parameters pertinent to musical creation. Beyond the commonly utilized EDA and EEG data, the dataset also encompasses evaluations by musicians of their own flow state and the musical audio. Notably, these streams of data have been synchronized along a timeline, permitting us to compare and analyze relationships between different types of data. This richness offers a macroscopic yet detailed perspective, not only aiding in revealing the connection between a musician's physiological state and the state of flow during musical creation but also deepening our understanding of how this state interplays with physiological responses and musical output.

## Embodied Musicking Dataset

"Embodied Musicking," central to the dataset's philosophy, also plays a pivotal role in our research. The dataset mirrors diverse musicians' improvisational performances of the same jazz piece, "How deep is the ocean," allowing us to observe how they harmonize their body with musical sound events. The nature of improvisation demands musicians to make decisions in split seconds, making this dataset an ideal platform to observe the interplay between body and mind in improvisational music.

The idea of "Embodied Musicking" stems from a cognitive perspective, proposing that musical creation isn't solely a cerebral activity but also involves bodily participation. In the musical realm, a musician's body serves as a vital medium for expression and communication. Viewed socially, it underscores the characteristic of musical creation as a social endeavor, necessitating collaboration and interaction between musicians. This interaction extends beyond the cognitive realm, encompassing coordination of the body and emotions. The multidimensional concept of "Embodied Musicking" aims to capture the interplay among cognition, the body, and the social aspects in musical creation, with the "Embodied Musicking Dataset" providing comprehensive data support.

In summary, our selection of this dataset is driven by the aspiration to understand the intricacies of musical creation from multiple angles and dimensions. We also hope to furnish potent references and insights for interdisciplinary research.

# Data Preprocessing

## Handling Missing and Outlier Values

Before diving deep into data analysis, it's paramount to ensure the quality and completeness of the data. We observed some missing values in the data of certain musicians. To address this, we employed interpolation-based methods to fill these gaps. However, in specific scenarios, such as when a missing value might reflect a genuine physiological or psychological state, we retained N/A values to avoid introducing unnecessary biases into the data analysis.

Given that physiological data, like EDA and EEG, can be influenced by external disturbances, equipment errors, or other noise, we meticulously cleaned the data. Specifically, we utilized statistical techniques and box plots to detect potential outliers. This was especially crucial as we identified consistent anomalous values in the data of a few musicians at the beginnings and ends of certain performances. For these cases, we undertook a detailed examination to determine whether these outliers reflected some actual physiological or psychological phenomena. Based on these analyses, we made informed decisions to ensure the reliability and accuracy of the data.

## Data Synchronization

In this project, data synchronization is pivotal, ensuring that different types of data (physiological, psychological, and audio) correspond accurately in a time series. This is vitally important for subsequent multidimensional analyses and pattern recognition.

Although the data was synchronously collected, ensuring that different physiological data items are aligned on a timeline, eliminating the need for additional physiological data synchronization steps. However, when delving into more profound time series analyses, we also needed to adjust data windows and granularity according to the characteristics of the musical performance. To address this, we employed a synchronization method based on the backing track audio file. Since the dataset used the same backing track audio file for each performance to maintain a consistent time framework during collection, this provided us with a reliable way to calibrate music bars and beats, achieving precise data synchronization. The specific steps are as follows:

1. Conduct musical analysis on the backing track, extracting crucial information, including the total duration of the music, BPM (Beats Per Minute), and the bars and beats corresponding to different timestamps.
2. Use this extracted information as a reference to segment and synchronize the data.

# Data Analysis

## Exploratory Data Analysis (EDA)

Before venturing into intricate data analyses, understanding the fundamental characteristics of the dataset is essential. EDA allows us to get an initial feel for the data, identifying potential patterns, outliers, or other features. This will be achieved through basic statistical methods (like mean, median, variance, etc.) to describe the data's central tendency and dispersion, and visualization tools (such as histograms, box plots, scatter plots, etc.) to display data distribution and relationships.

## Correlation Analysis

To further explore whether there exists a linear or monotonic relationship between physiological parameters (e.g., EDA, EEG) and psychological states (e.g., flow state), correlation analysis was chosen as the method. We used the Pearson correlation coefficient to test for linear relationships between variables and the Spearman correlation coefficient to test for monotonic relationships. Moreover, considering the potential multi-time scale relationship between improvised music and physiological responses, we also introduced localized correlation analysis.

Beyond basic correlation coefficients, we also considered multi-scale and multi-level analyses. For instance, within different time windows, we contemplated using localized correlation analysis to better capture short-term dynamic changes. This is because improvised music and physiological responses might have different association patterns at different time scales.

## Time Series Analysis and Machine Learning

Given the complexity of the music creation process, a single analytical strategy might be insufficient to capture all patterns and relationships. Therefore, we integrated time series analysis with machine learning methods. For example, through K-means clustering, we can segment data and explore physiological and psychological response patterns under different time windows.

## Statistical Software and Libraries

In this research, we adopted Python as the core programming and analysis tool due to its robust capabilities and extensive applications. Below are the primary libraries we utilized and a brief introduction to their functions:
1. Pandas:
    This is the most popular data processing and analysis library in Python. We used Pandas

for data import, cleaning, transformation, and basic statistical analyses. Its efficient DataFrame structure makes large-scale data operations intuitive and efficient.

2. SciPy:
SciPy is a potent scientific computing library, offering a plethora of mathematical algorithms and functions. In this research, we used SciPy for advanced statistical tests, signal processing, and other intricate tasks.

3. Matplotlib & Seaborn:
These two libraries are our go-to for data visualization. Matplotlib provides fundamental plotting capabilities, while Seaborn offers a high-level interface for statistical data visualization, simplifying the creation of complex charts.

4. NumPy:
Being the cornerstone of scientific computing in Python, we used NumPy for numerical computations, like linear algebra and Fourier analysis. Its array structure also underpins several other libraries (e.g., Pandas and SciPy).

5. scikit-learn:
When it comes to machine learning algorithms, we turned to scikit-learn. This is a comprehensive machine learning library, offering a vast array of both simple and advanced algorithms, from regression and classification to clustering.

6. statsmodels:
For more intricate statistical model analyses, we utilized the statsmodels library. It offers implementations of various statistical models, from linear models to time series analysis.

The choice of Python and its rich libraries as our analysis tools was based on their power, flexibility, and community support. As the research progresses, we may introduce more specialized tools and libraries as needed. In summary, Python provides us with a comprehensive, cohesive environment, supporting the entire process from data preprocessing to advanced analysis.

# Results

## Experience 1: Global Overview

In this phase, our primary focus was on obtaining a holistic understanding of the dataset by utilizing various types of data, such as EDA, EEG, and skeleton data. Our intention was not to draw direct inferences but to identify areas that might exhibit significant correlations, guiding our subsequent investigations. This stage aimed to provide a global view of the Embodied Musicking dataset, capturing essential patterns and anomalies present throughout. The code associated with this phase can be referred to in Appendix 1.

We selected 25 json files from the dataset, each representing the improvised performance data of a musician in response to a consistent backing track.

## Data Integrity Check

A preliminary check on data integrity revealed that most columns were devoid of missing values, indicating high data integrity for these features. Notably, the "flow" column had about 0.59% missing values, which is of particular concern for our study. As we move to later stages of data processing, strategies like imputation or sensitivity analysis might be needed to address these absent values. Having a near-complete dataset is imperative, especially in a multidimensional data study like this where the goal is to interconnect different types of data. Missing values, especially in key variables like "flow", can introduce biases or diminish the validity of statistical tests. For the sake of brevity in this phase, we employed median imputation.

## Basic Descriptive Statistics

The dataset comprises 52 features. Based on their names, they can be broadly categorized into two groups: synchronization information, which includes attributes like session id, chorus id, and backing track position; and collected data, which encompasses EDA, EEG, skeletal motion data, and self-reported flow data from the musicians.

Some numerical insights include:

1. Sync Delta: With an average of 30.03, the substantial standard deviation of 4247.18 suggests significant data variability. It seems there might be outliers or extreme values, as indicated by the minimum value of -331500 and a maximum of 3945.
2. Hardware Bitalino (EDA): The average is around 292.49 with a standard deviation of 175.77, suggesting a moderate range of distribution around the mean. The range from a minimum of 19 to a maximum of 1020 implies potential outliers or extremes.
3. Hardware Brainbit EEG (T3, T4, O1, O2): The means range from 23,413 to 93,129, but with a high standard deviation, indicating these features have substantial variability.
4. Hardware skeleton nose(x, y): A brief look at the skeletal data for the nose showed averages of 245.68 (for x-coordinate) and 123.44 (for y-coordinate). However, both x and y have a minimum value of -1, indicating possible missing or inaccurate values.

It's noteworthy that the EDA and EEG data have differing units and scales, highlighting the need for standardization or normalization in subsequent stages. High standard deviations in certain features, like sync_delta and EEG parameters, point to variability but also warrant attention towards potential outliers, which will be addressed during data cleansing.

## Initial Data Visualization

One of the objectives of this phase was to visualize the distribution of key features through histograms. For the Hardware Bitalino (EDA), the histogram displayed a right-skewed distribution, suggesting that lower EDA levels are more prevalent throughout performances. Similarly, Hardware Brainbit EEG (T3 and T4) also exhibited a right-skewed distribution, with most values concentrated on the lower end. In contrast, the distribution of "flow" values appeared relatively uniform but with certain peaks, indicating variations in self-reported flow states.

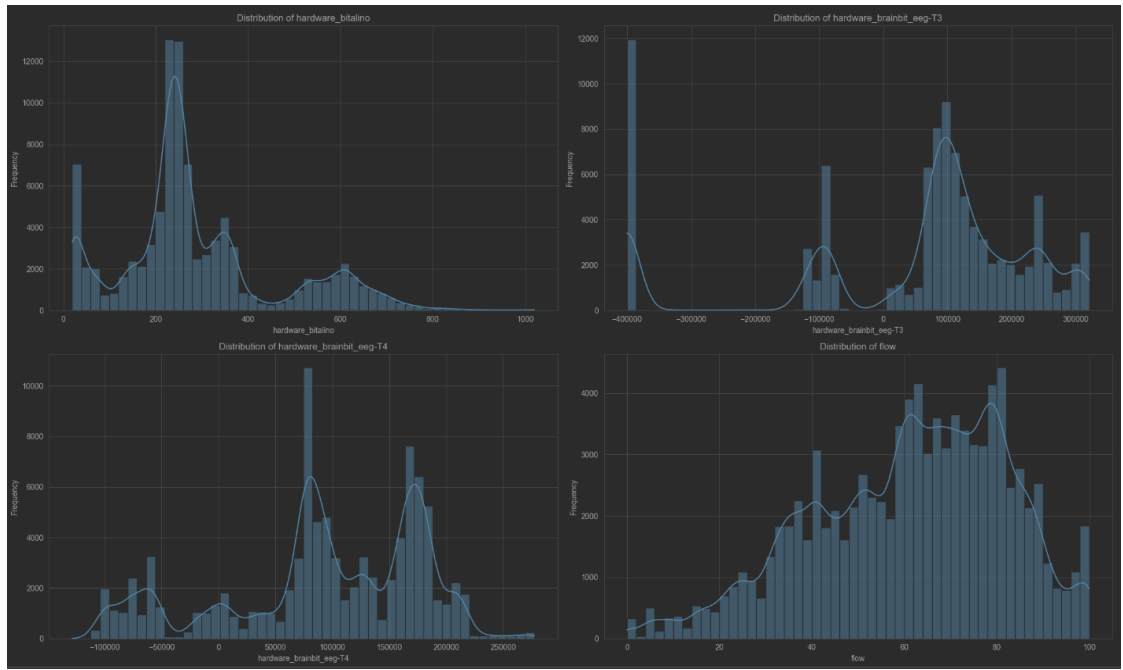

Figure 1. Histograms of Flow, EEG-T3, EEG-T4, and EDA

Correlation Heatmap
A heatmap was generated to visualize the correlation between these variables. The EEG channels (T3, T4, O1, O2) exhibited strong inter-correlations, which is expected since they all measure electrical activity in the brain. The "flow" column did not show strong correlations with physiological datapoints like EDA or EEG, possibly suggesting the self-reported "flow" states by musicians might not be as reliable, or it represents a more intricate structure that can't be directly gauged through simple physiological metrics. Skeletal data, specifically X and Y coordinates for body parts like the nose, neck, and shoulders, showed significant correlations, implying that these body parts might move synchronously during performances.

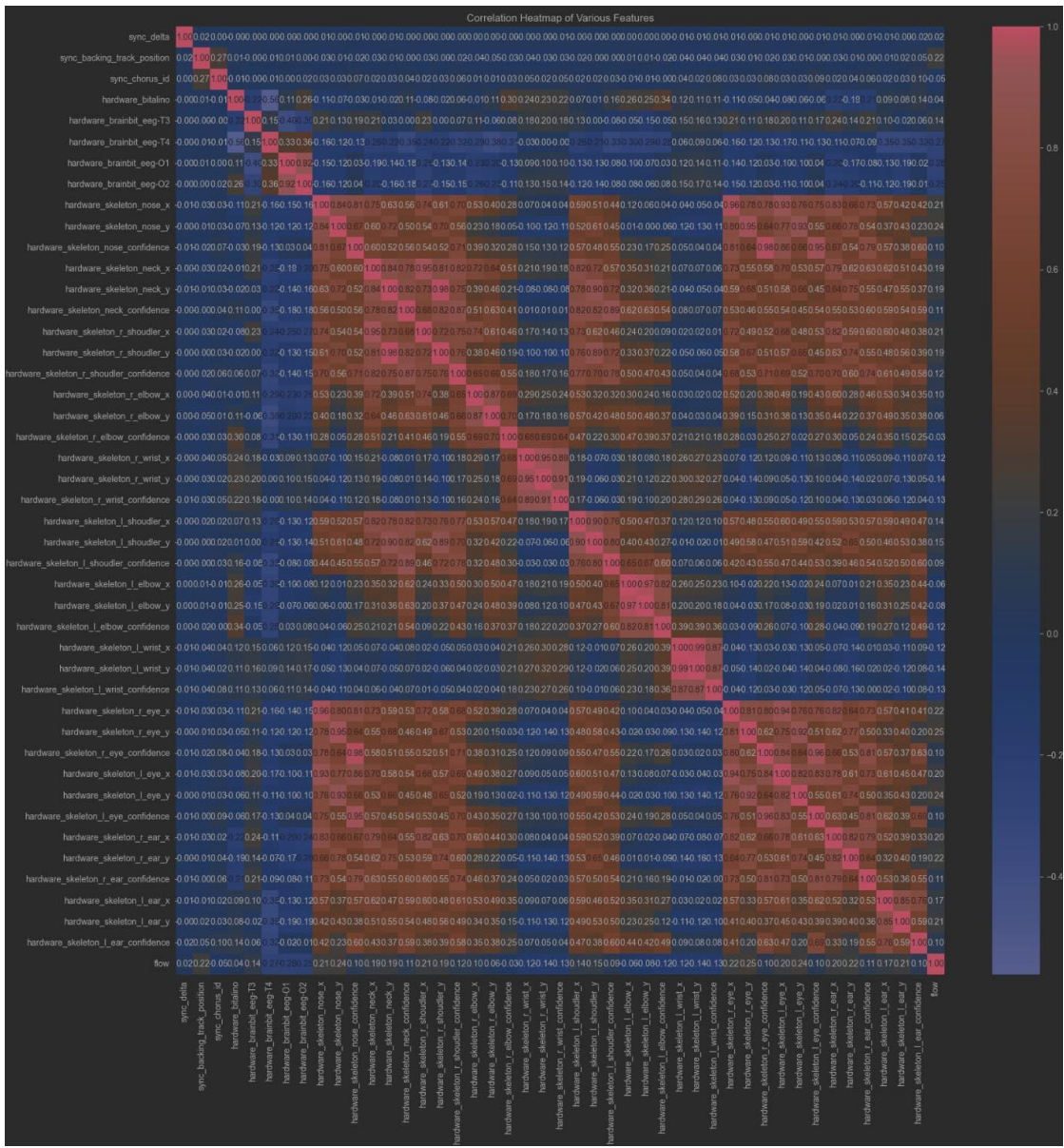
Figure 2. Correlation Heatmap

The lack of robust correlations between "flow" and physiological variables indicates that a deeper, more intricate analysis is required to uncover any underlying relationships. Nonetheless, this heatmap serves as a foundational step in understanding the multidimensional nature of the dataset.

## Experience 2: Deep Dive into a Single File

The objective of this phase is to delve into the data characteristics of a single performer, laying the groundwork for subsequent analyses across multiple files. Choosing which musician to deeply analyze is a pivotal decision. To ensure the reliability and representativeness of our analysis, we established the following criteria:

1. **Data Integrity**: We prioritize musicians whose data showcases the highest integrity, meaning the least amount of missing values. This provides us with a more continuous and comprehensive time-series data, enhancing the precision of our analysis.
2. **Data Representativeness**: We opt for musicians whose physiological data and "flow" scores closely mirror the average values of the entire dataset, ensuring that our analysis yields generalizable results.
3. **Data Variability**: Musicians who exhibit significant variations in their physiological data and "flow" scores during performances are selected. This enhances our ability to discern possible correlations and patterns.

Given the aforementioned criteria, we decided to delve into the data of musician "Jn3VvBWcnDESzN9gUTh3bN". This musician boasts high data integrity, a "flow" score with a minimal deviation from the dataset's average, and notable variations across different performances.

## Data Integrity Check

- Missing Values: The data is largely intact, with only 1 and 3 missing values detected in the sync_delta and flow attributes, respectively.
- Outliers: We employed the Interquartile Range (IQR) method to detect outliers across multiple columns. These outliers can be further examined in Table 1, which only documents attributes with non-zero missing or outlier values. The complete data can be reviewed through the code in Appendix 2. We observed that the majority of these outliers are predominantly concentrated within the skeletal data. Such anomalies might arise due to various reasons: issues with the equipment or sensors during data collection, external interferences leading to erroneous readings, or potentially rapid, atypical movements made by the musician. Interestingly, the confidence levels of some body parts align with their corresponding outlier values, while others exhibit significant discrepancies. Consequently, based on the distinct attributes of these outliers, we need to decide how to address them in our detailed analysis.

| Data items | Missing values | outliers |
| --- | --- | --- |
| sync_delta | 1 | 4 |
| sync_chorus_id | 0 | 59 |
| flow | 3 | 0 |
| hardware_skeleton_l_ear_confidence | 0 | 36 |
| hardware_skeleton_l_ear_x | 0 | 572 |
| hardware_skeleton_l_eye_confidence | 0 | 289 |
| hardware_skeleton_l_eye_x | 0 | 296 |
| hardware_skeleton_l_shoudler_x | 0 | 252 |
| hardware_skeleton_l_shoudler_y | 0 | 283 |
| hardware_skeleton_l_wrist_confidence | 0 | 230 |
| hardware_skeleton_l_wrist_x | 0 | 230 |
| hardware_skeleton_l_wrist_y | 0 | 230 |
| hardware_skeleton_neck_confidence | 0 | 93 |
| hardware_skeleton_neck_x | 0 | 118 |

| hardware_skeleton_neck_y | 0 | 137 |
| hardware_skeleton_nose_confidence | 0 | 349 |
| hardware_skeleton_nose_x | 0 | 269 |
| hardware_skeleton_r_ear_x | 0 | 287 |
| hardware_skeleton_r_ear_y | 0 | 283 |
| hardware_skeleton_r_elbow_confidence | 0 | 175 |
| hardware_skeleton_r_elbow_x | 0 | 105 |
| hardware_skeleton_r_elbow_y | 0 | 171 |
| hardware_skeleton_r_eye_confidence | 0 | 327 |
| hardware_skeleton_r_eye_x | 0 | 302 |
| hardware_skeleton_r_shoudler_confidence | 0 | 99 |
| hardware_skeleton_r_shoudler_x | 0 | 113 |
| hardware_skeleton_r_shoudler_y | 0 | 121 |
| hardware_skeleton_r_wrist_confidence | 0 | 71 |
| hardware_skeleton_r_wrist_x | 0 | 414 |
| hardware_skeleton_r_wrist_y | 0 | 582 |

Table1. Missing Values and Outliers

## Synchronization Attributes Analysis

### Backing Track Position & Delta

The 'Backing track position' acts as a continuously increasing sequence, functioning as a timestamp or positional identifier for each consecutive recording in a file. On the other hand, 'Delta' captures the time interval or positional change between two successive data records. Within the scope of time-series data, 'Delta' typically represents a difference or variation. However, in this particular context, it specifically denotes a temporal difference. This understanding can be solidified by calculating the difference between two successive data points of the 'Backing track position'. Upon wrapping up this section of the analysis, we can unearth a critically significant attribute of the dataset: the sampling rate.

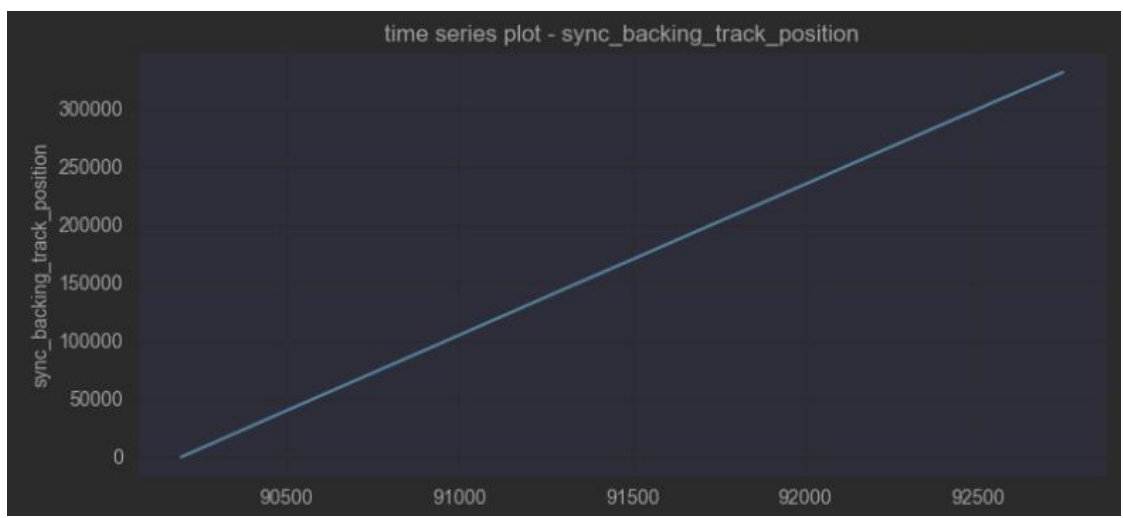

Figure 4. Backing track position time series chart

**Data Distribution**

Most Delta values cluster around the 125-150 interval, signifying that the majority of data recording intervals remain relatively stable.

**Outliers**

We employed the Interquartile Range (IQR) method to detect outliers across multiple columns. These outliers can be further examined in Table 1, which only documents attributes with non-zero missing or outlier values. The complete data can be reviewed through the code in Appendix 2. We observed that the majority of these outliers are predominantly concentrated within the skeletal data. Such anomalies might arise due to various reasons: issues with the equipment or sensors during data collection, external interferences leading to erroneous readings, or potentially rapid, atypical movements made by the musician. Interestingly, the confidence levels of some body parts align with their corresponding outlier values, while others exhibit significant discrepancies. Consequently, based on the distinct attributes of these outliers, we need to decide how to address them in our detailed analysis.

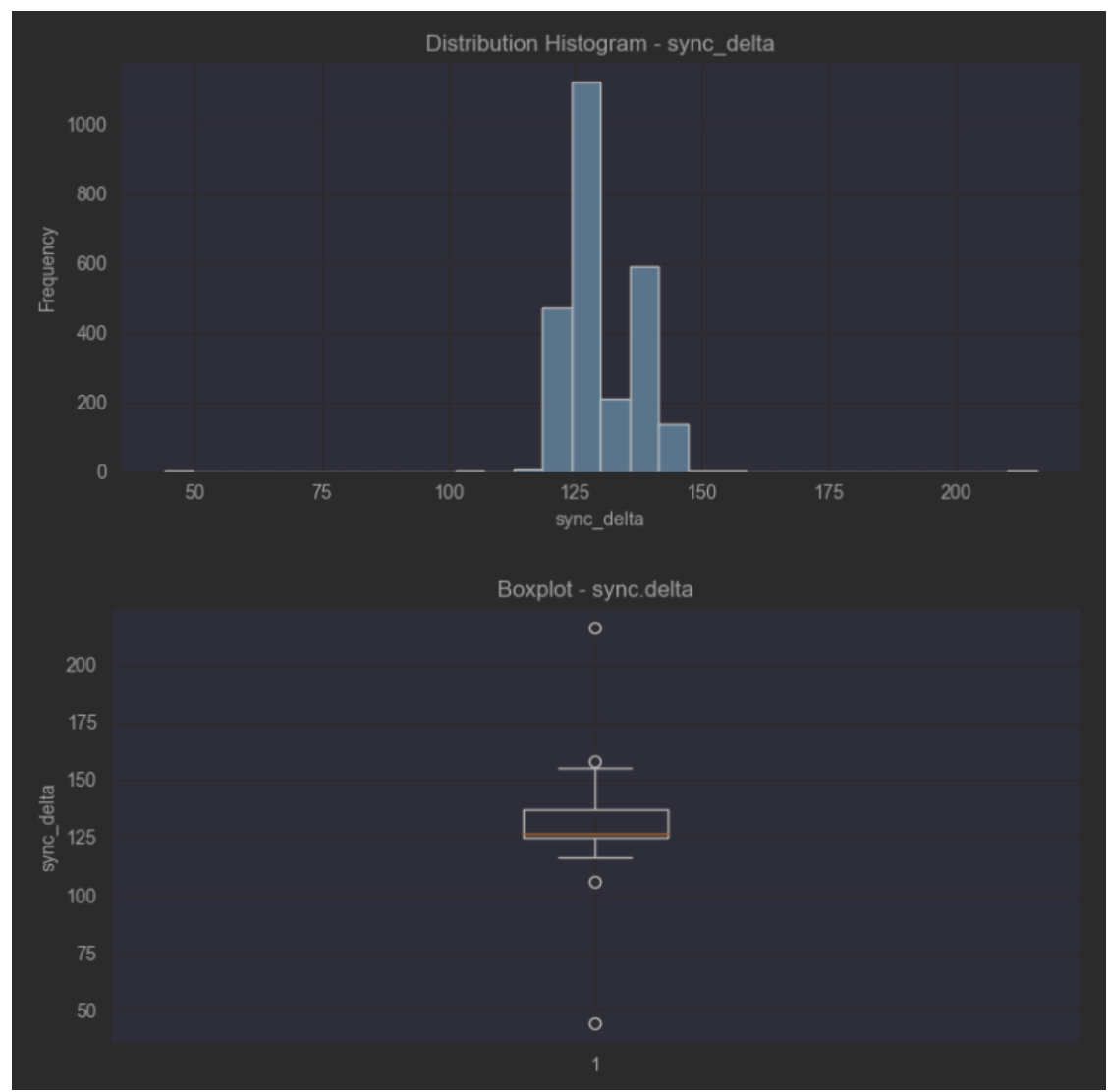

Figure 5. Histogram and Box Plot of Delta

**Temporal Stability of the Time Series**:

From our visualization of the time series, it's evident that the 'delta' exhibits relative stability over time. However, periodic fluctuations were also noted, hinting at potential transient changes or interruptions during the recording phase.

From a practical standpoint, the 'delta' offers invaluable insights into the frequency of data recording. A sudden surge in the 'delta' value may imply a deceleration in data recording or perhaps an interruption during a specific time frame. Conversely, a sudden dip in the 'delta' might indicate an acceleration in the data recording speed. In essence, 'delta' furnishes us with critical information about temporal variations during the data collection process, potentially aiding in understanding shifts in other attributes and ensuring data coherence and integrity.

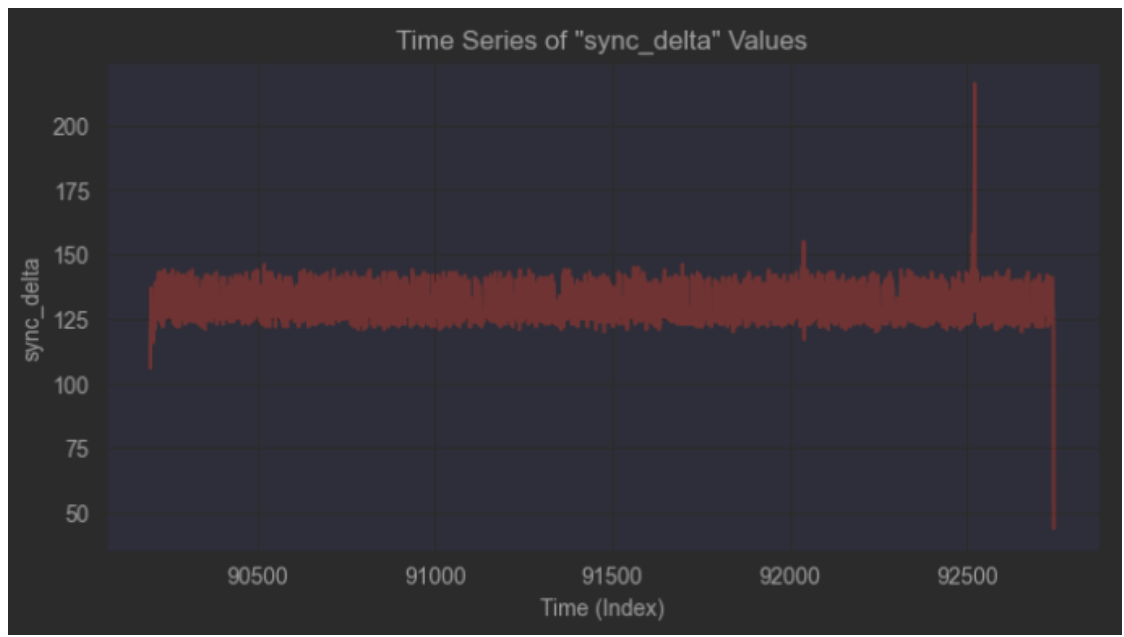

Figure 6. Time Series of 'sync_delta' Values

**Sampling Rate**:

To determine the sampling rate, it's essential to compute the continuous differences in the 'backing track position', thereby ascertaining the time interval between data points. Some slightly larger discrepancies were noted, like the value 216. Such variations might stem from the instability of the data collection device but are still within the acceptable fluctuation bracket. While similar outcomes could be achieved using 'delta', and the statistical values closely mirror those of 'backing track position', we opted to use the latter for calculations due to its absence of outliers.

The sampling rate (Hz) can be deduced using the formula: Sampling Rate = 1/Average Sampling Interval (seconds). For this particular musician, the average sampling interval stood at 130.153121ms, yielding a sampling rate of approximately 7.683257937395139Hz. This translates to roughly 7.68 samples per second.

Such a sampling rate implies that we can only glean around 8 data points every second. Based on the Nyquist theorem, our analysis is restricted to frequency components below 3.84Hz.

This indicates our potential inability to accurately analyze any signal that surpasses this frequency. For rapidly evolving physiological or psychological events, this rate might be inadequate to capture all intricacies. However, a lower sampling rate could also render the data smoother, as high-frequency noise or variations might either go unrecorded or some high-frequency noise might blend into the low-frequency signals. This underscores the impending need to tailor our analysis strategies and data based on the sampling rate.

**Chorus id**

The chorus id is a pivotal attribute within our project. In a standard musical context, a "chorus" refers to the repeating section of a song. However, in this setting, it indicates distinct playthroughs of the same song by the musician. Specifically, the chorus id consists of seven unique values, enabling us to segment each musician's performance into seven distinct parts. Values 1-5 represent the musician's five individual playthroughs of the song. Meanwhile, 0 and 999 likely signify the beginning and end of the performance, respectively. It's noteworthy that the count of data points for 999 aligns precisely with the number of anomalies we identified earlier.

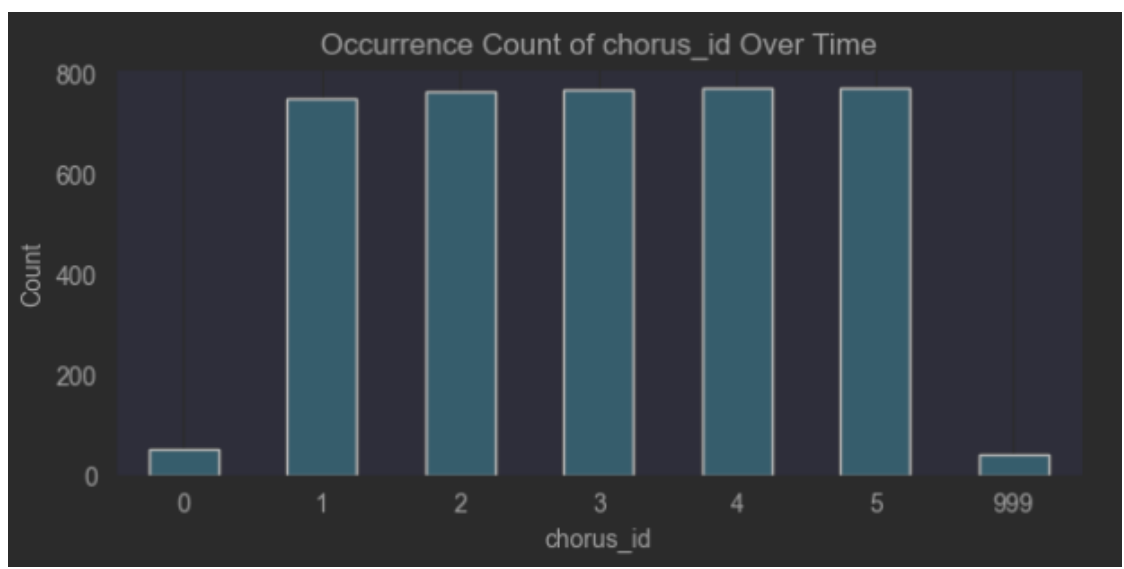

Figure 7. Distribution of Chorus id over time series

## Analysis of Physiological Data and Flow

**Flow**

Flow is one of the core attributes of our project and forms a part of our objectives. In this phase of exploration, it's essential to integrate flow with other data items to preliminarily investigate the presence of specific patterns.

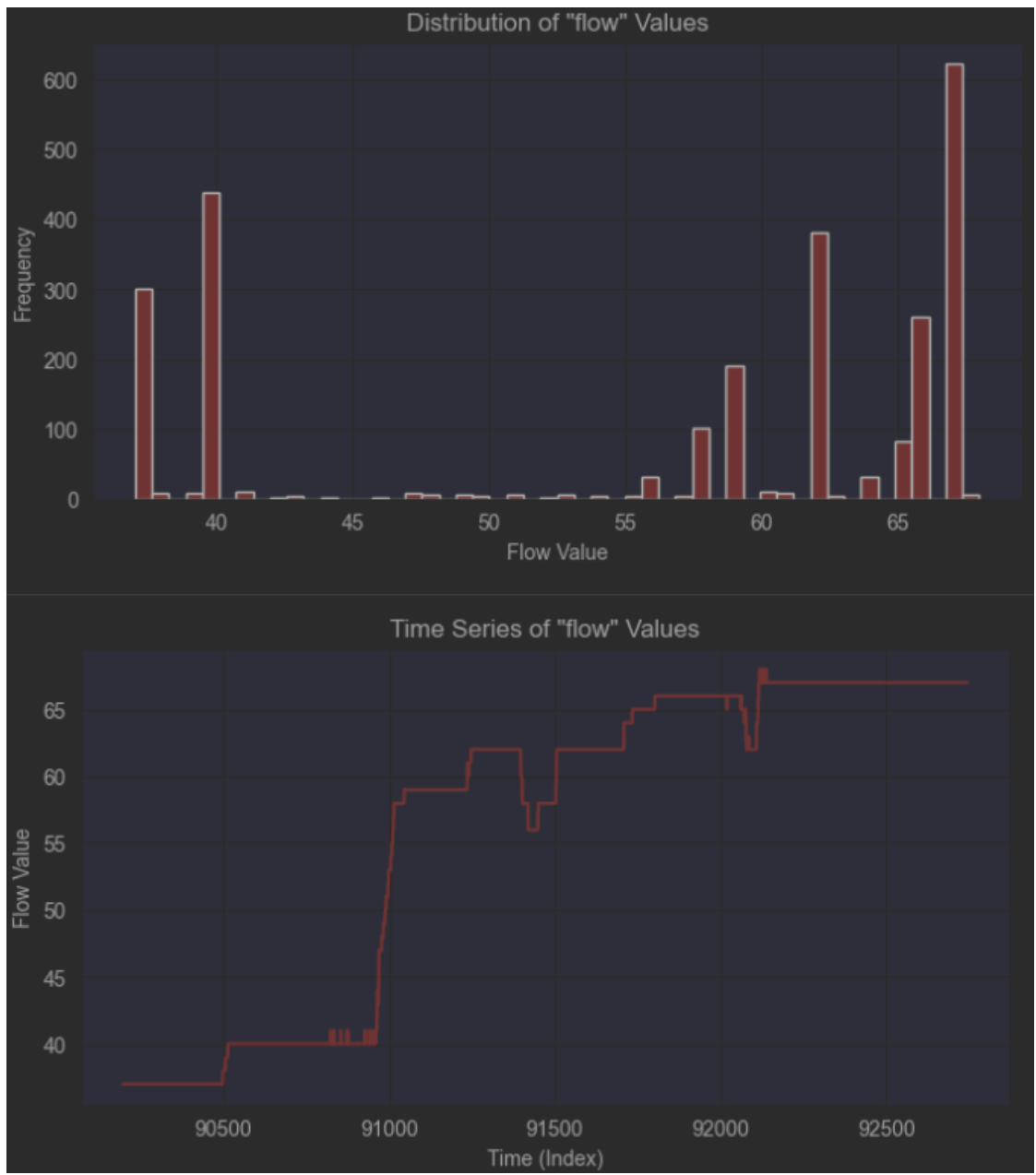

Figure 8. Bar Chart and Time Series Plot of Flow

**Data Distribution**:

The values for flow range between 37 and 68. Notably, certain flow values, such as 37, 40, 58, 59, 62, 66, and 67, have a noticeably higher count compared to others, with the flow value of 67 having the most data points, totaling 623. Some flow values have a minimal count, for instance, data points with flow values ranging between 42-56 are all less than 10. The average value stands at 56, and the median is 62, placing it at a relatively higher level within the distribution range. The time series plot reveals an ascending trend in flow values over time, starting at lower values but gradually increasing as time progresses.

These observations suggest that the musician had varying levels of self-assessment or flow experiences across different improvisational phases. The musician seems to be more engaged during the middle and later stages of the performance, potentially resulting in music that resonates more emotionally and showcases greater technical proficiency. Another point worth noting is that the "flow" state is subjective and self-reported, making it susceptible to biases such as social expectations and recall bias. Consequently, while the data indicates a high degree of engagement, caution is required when interpreting this metric in subsequent analyses.

**EDA (Electrodermal Activity)**

Electrodermal Activity (EDA) is among the pivotal indicators in our analysis. The distribution range for EDA data spans from 318 to 649, with an average value of 467.44. A majority of data points are concentrated around the 490 mark.

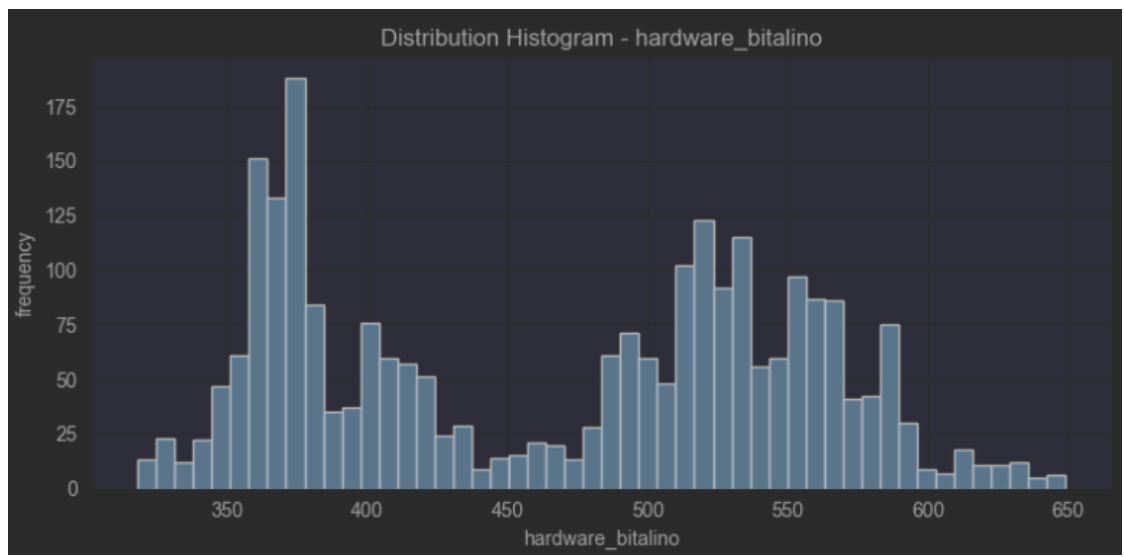

Figure 9. Histogram of EDA

**Time Series and Moving Average**:
By computing the moving average for EDA data, we can mitigate short-term fluctuations and noise, offering a clearer view of the overall EDA trend. The blue line in the graph represents the raw EDA data, while the red line illustrates the moving average taken every 10 seconds. This moving average elucidates the general trend in EDA data without the distraction of short-term variations. We can discern an upward trajectory in EDA throughout the performance, especially in the latter stages. This could indicate that as the performance unfolds, the musician might be experiencing heightened physiological arousal or motivation.

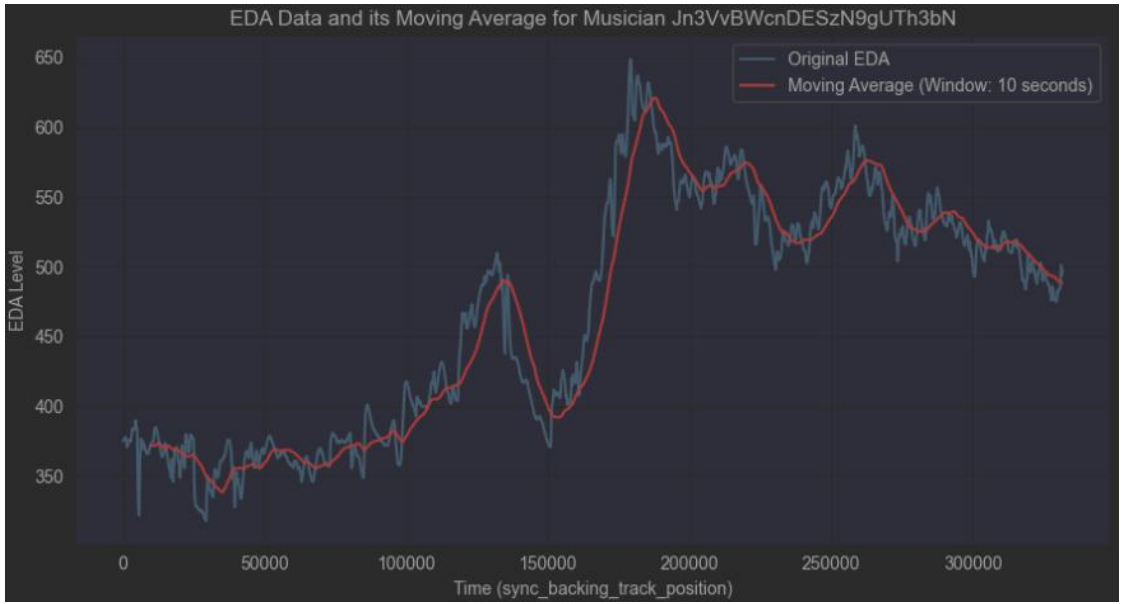

Figure 10. Time Series and Moving Average Plot for EDA

**Peak Detection**:

Peaks in the EDA signal might correspond to physiological or emotional "responses" experienced by the musician. For instance, when a musician feels anxiety, excitement, or undergoes a certain emotional climax, their EDA might exhibit transient peaks. Consequently, we plotted EDA data alongside "flow" values to exhibit peaks in conjunction with the reported flow states.

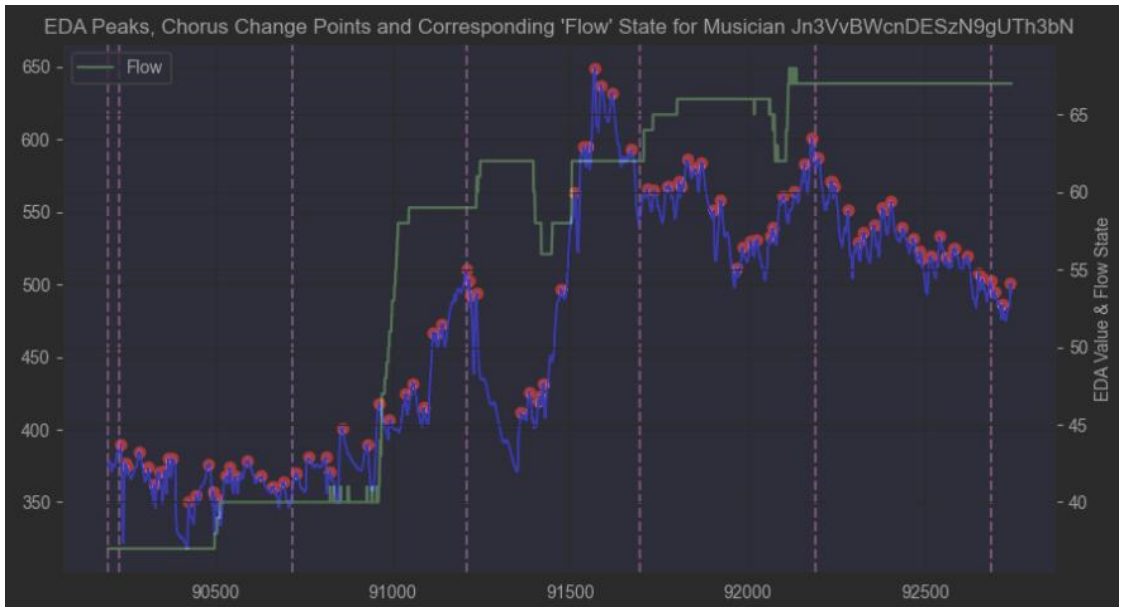

Figure 11. Time Series Plot with Marked EDA Peaks

In the chart, the blue line signifies the EDA values, with red dots indicating detected significant peaks. The green line represents the reported flow state by the musician, while pink dashed lines segment the different choruses based on the chorus id. Under the delineation of chorus id, no discernible pattern emerges across the five choruses, suggesting a need for a more granular segmentation in future analyses. The trends of EDA and flow exhibit some parallels; at certain points where EDA peaks are evident, fluctuations in the flow state are also apparent, and occasionally, there seems to be a temporal lag between the peaks of the two. It's imperative to acknowledge that EDA is a sensitive metric, potentially influenced by a myriad of factors like ambient temperature, humidity, and individual skin conductivity. Hence, even though the data range indicates stability, EDA peaks don't always align with high or low "flow" states and can't be conclusively deemed as a comprehensive reflection of the musician's psychological state. Yet, on the flip side, while EDA might not always directly correlate with flow states — given that flow is a self-reported metric and might not be as sensitive or detailed as physiological indicators like EDA — EDA could capture transient changes the musician undergoes during a performance, changes that might elude the musician's own perception. Thus, there remains a plausible link between EDA and the musician's psychological state, a connection we aim to substantiate in subsequent multi-file analyses.

### Electroencephalogram (EEG)

Though constrained by the frequency resolution, which only captures signals below 3.9 Hz, we might miss out on some frequency bands. This makes it challenging to conduct meaningful analysis based on EEG frequencies. Nevertheless, we can attempt to understand brain activity over time through time-domain analysis.

**Data Distribution**:
All four EEG attributes exhibit a bimodal distribution, though the bimodal nature of EEG-T3 appears relatively indistinct.

**Trend Analysis**:
EEG data is often susceptible to noise and may also be influenced by non-cognitive factors such as muscle movements, eye blinks, or even electrical interference. Hence, a moving window average was employed during the trend analysis to minimize potential noise interference, making the data more coherent and assisting us in observing the primary signal trends more clearly. The window size was set to 10 seconds, based on the previously determined 7.68Hz sampling rate. Upon observation, we discern that the different EEG channels (T3, T4, O1, O2) exhibit highly similar patterns throughout the entire time span. The high correlation among the four channels is evident even without a formal correlation analysis.

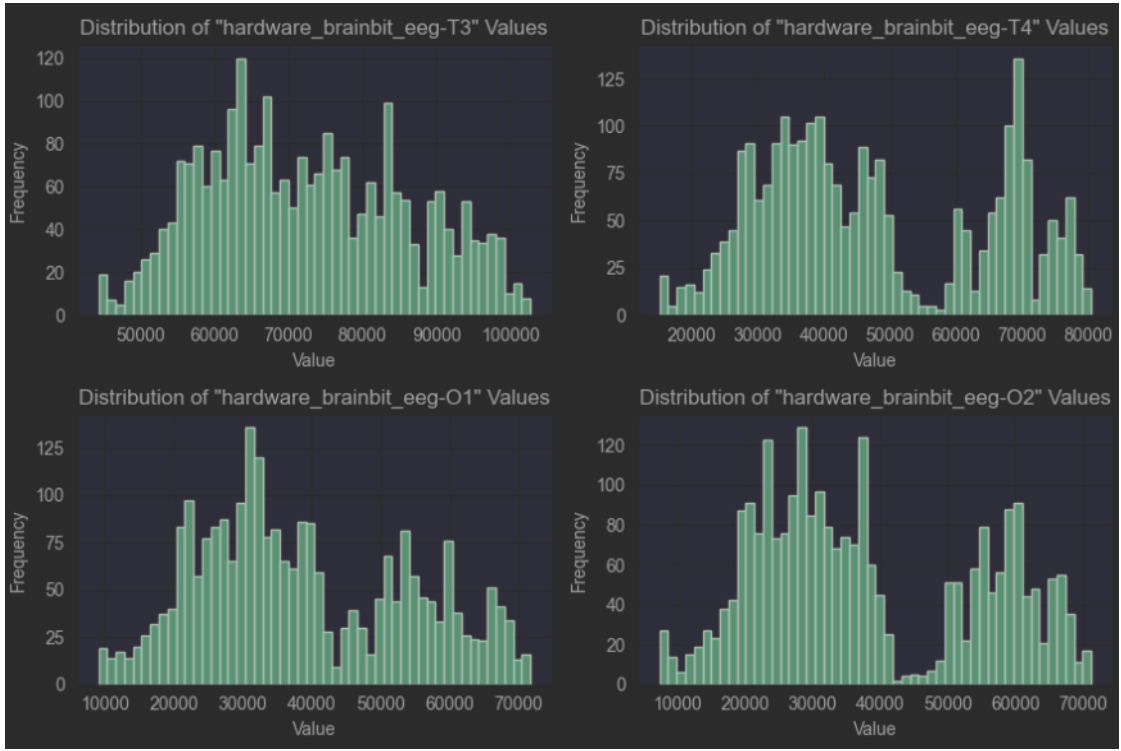
Figure 12. Histogram of the Four EEG Channels

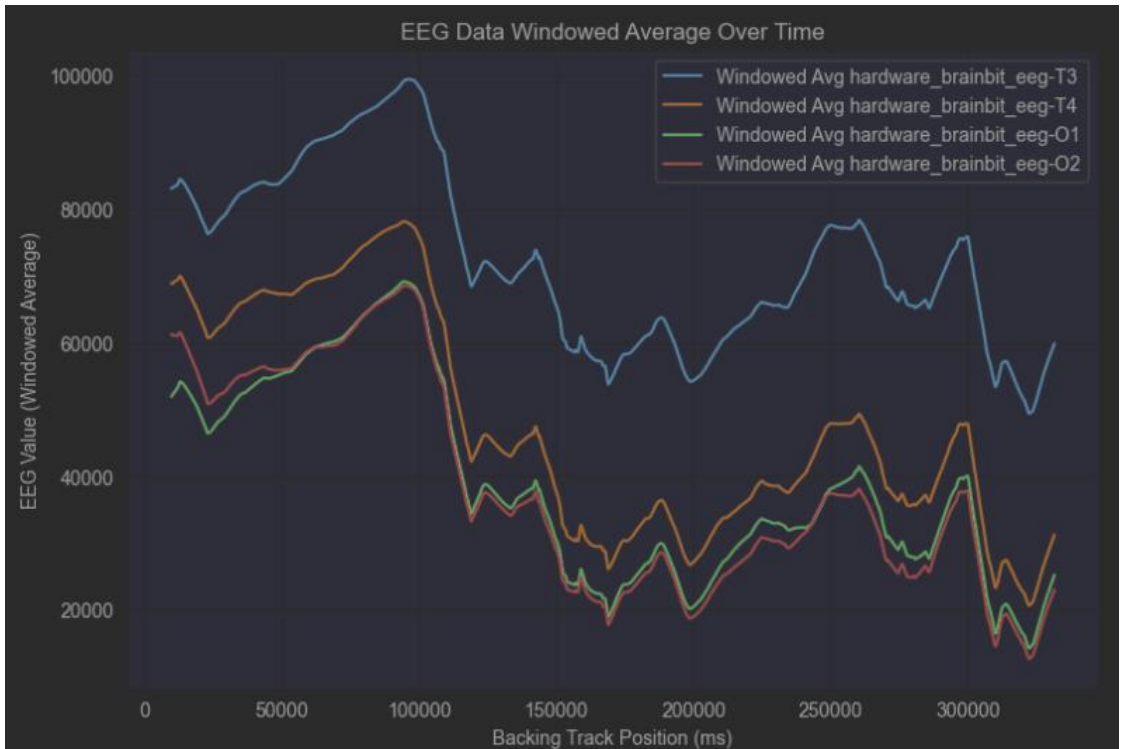
Figure 13. Moving Window Average Time Series for the Four EEG Channels

**Time-Varying Variance**:
The variance of EEG signals might fluctuate within specific time intervals, aiding our understanding of the stability or variability of brain activity. All EEG channels throughout the time span depict almost synchronized variance alterations. This suggests that at certain moments, the EEG signal's volatility or instability might escalate. Peaks in variance could mirror changes in the musician's brain activity triggered by specific activities or states during the performance. In certain timeframes, the variance across all channels is comparatively low, suggesting relative stability in brain activity during these periods.

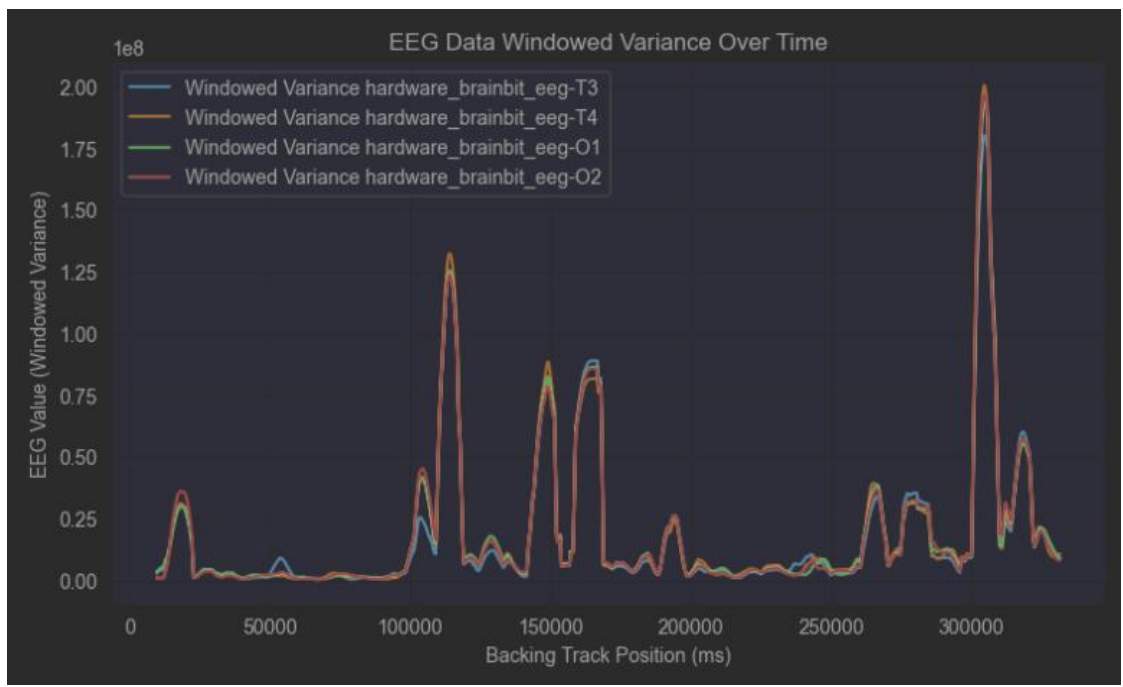

Figure 14. Time-Varying Variance Time Series for the Four EEG Channels

From the three analytical methods above, it's evident that all four channels consistently mirror each other across various metrics. Such consistency is generally unusual since these channels represent different functional areas of the brain. Considering the sampling rate data, we might have lost some crucial high-frequency information distinct in different brain areas. Intriguingly, contrasting the progressively increasing trend of flow, the EEG generally shows a declining trajectory. We could speculate that as the performance progresses, musicians become more adept and confident. This proficiency and confidence might be mirrored in their Flow data, but manifest as a reduction in some brain region activities in the EEG data. Though this possibility exists, our primary focus needs to shift to other data items. EEG data with a 7.68 Hz sampling rate might not be comprehensive enough to capture all essential activities in the brain, especially pertinent α, β, and γ wavebands. This restricts our capability to extract meaningful insights from the data. Also, due to the high synchronization of the EEG, it becomes challenging to extract unique and differential information from these four channels. We'll delve deeper into this decision in the subsequent discussion section.

**Motion Data**

**Overall Activity Level**:
Before delving deep, due to the abundance of skeletal data, we first aim to get a rough understanding of its potential implications. We wish to take a holistic view of the musician's overall dynamism throughout the performance. By computing the average X and Y coordinates for all skeletal parts, we're granted a glimpse into the musician's overall movement activity. By simply observing the variations in average coordinates, we can quickly grasp the musician's movement trends. Averaging data across all parts can reduce noise, making overarching trends more pronounced. However, the downside is that blending all features might muddle the movement details of individual parts. Unaddressed outliers could also distort the visualization. At this juncture, we seek only a preliminary overview, checking if there are distinct active or passive phases, possibly offering intriguing directions for deeper analyses.

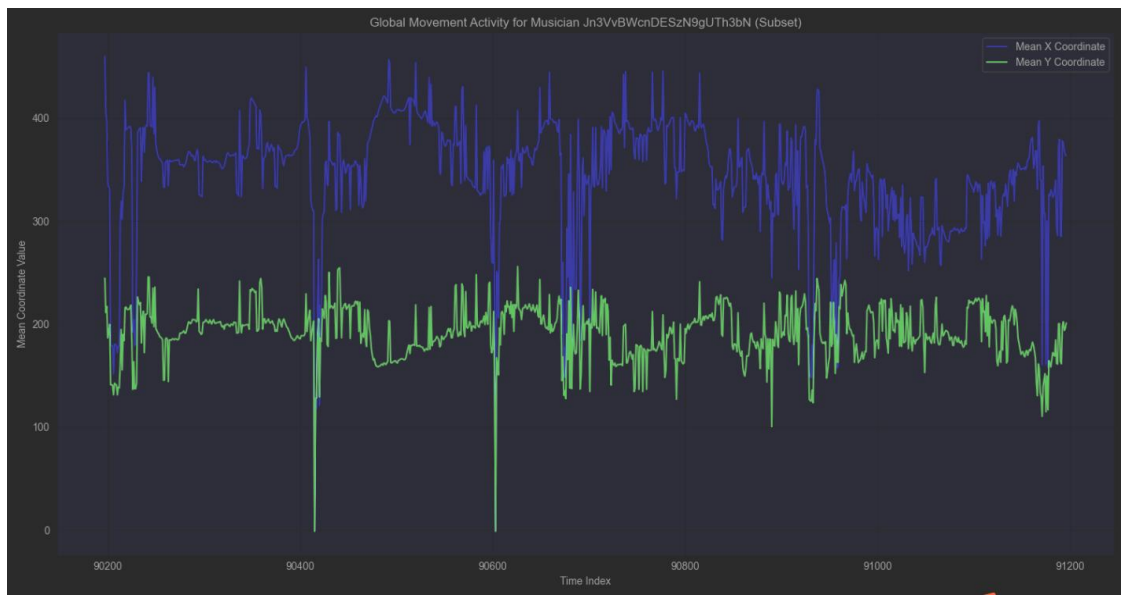

Figure 15. Overall Activity Level for All Skeletal Data

From the above chart, we observe: Both Mean X and Mean Y coordinates exhibit pronounced fluctuations, indicating significant movement by the musician during their performance. The trends for Mean X and Mean Y are largely analogous, hinting that the musician's movement might be chiefly in a specific direction, possibly forwards-backwards or left-right. We can pinpoint some peaks and troughs, possibly representing distinct moves or shifts.

**Movement Trajectory**:
Given that it's motion data, observing the trajectory and patterns of movement might bring new insights to our exploration. We selected three representative parts to attempt plotting trajectory and heat maps. We picked the left ear, left wrist, and right wrist, symbolizing movements of the head and upper limbs, respectively. The movement of the head is relatively independent, while the movements of the two hands might be interconnected.

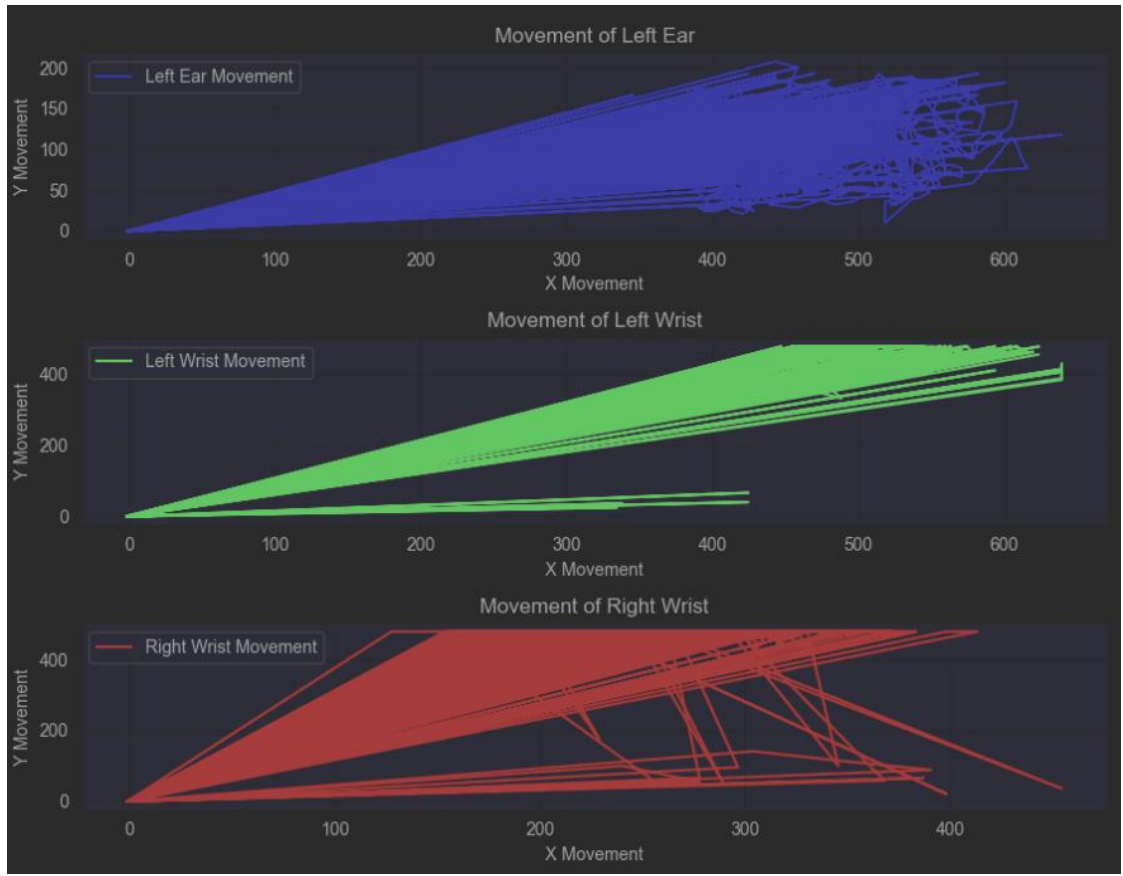

Figure 16. Trajectory of the Left Wrist on the x, y Plane

The trajectory chart showcases the continuous movement path of the musician's left ear, left wrist, and right wrist throughout the improvisation. This grants a dynamic view of how specific parts of the musician's body moved during the creative process. These trajectories offer a visual insight into the motion patterns and paths of body parts. For the parts observed simultaneously (both hands), if their paths mirror or run parallel in some segments, it might indicate coordinated or responsive actions between the hands. Variability in trajectories of different parts could link to the musician's expressiveness in the performance.

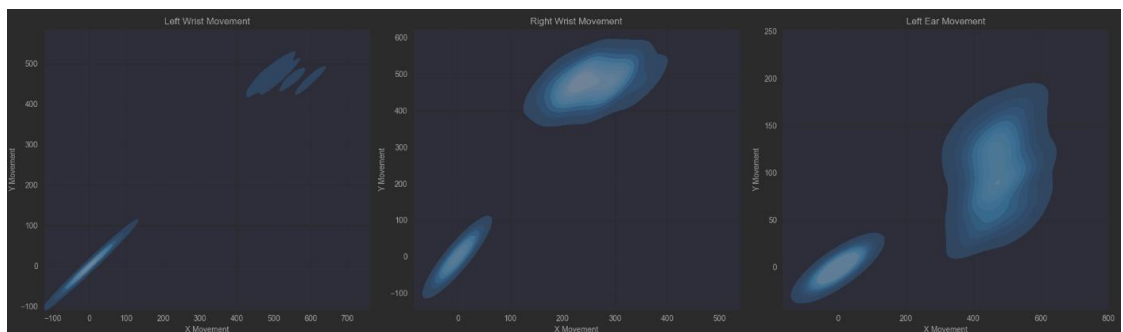

Figure 17. Heatmaps for Left Wrist, Right Wrist, and Left Ear

**Outlier Handling**:
Having briefly observed potential patterns in motion data, we need to decide on the treatment of outliers before a detailed analysis, as most missing and outlier values are concentrated in skeletal data.

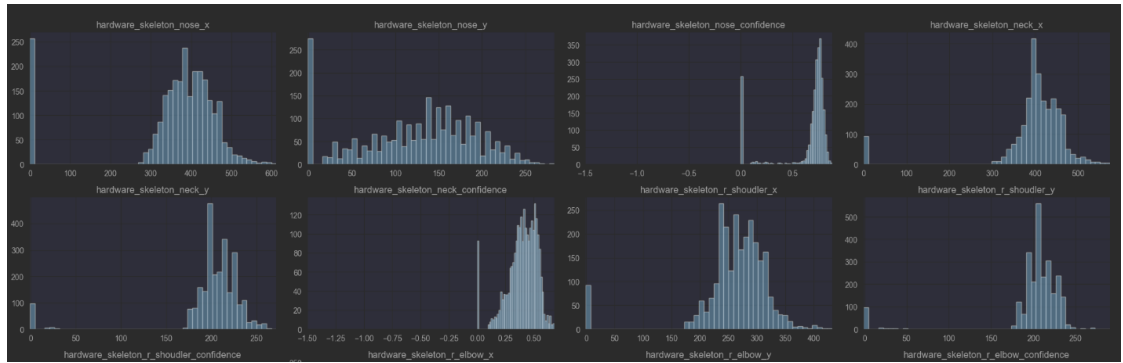

Figure 18. Histogram for Some Skeletal Data

After visualizing the distribution for most motion parts, we discern that almost all data exhibits a certain number of zeros. To validate this, we examined the raw data and found consistent zeros and -1s across all data points, possibly indicating issues during data collection, generating a plethora of unreliable data. Therefore, we counted rows for three scenarios: 1. Rows with data value -1. 2. Rows with data value 0. 3. Rows with confidence below 0.5. Figure 19 captures a snippet of the results, with full details available in Appendix 2's code. The results reveal a substantial portion of data with low confidence, and a high proportion of -1s, explaining why our motion trajectory visualization frequently reverts to zero. In light of this, we'll extract data with relatively higher confidence, exploring methods to render the data usable. Thus, we filtered attributes where the number of rows with -1s, 0s, or confidence below 0.5 are all less than 400, obtaining the columns shown in Figure 20.

| | Rows with -1 | Rows with 0 | Rows with Confidence < 0.5 |
|---|---|---|---|
| hardware_skeleton_r_elbow_y | 102 | 0 | 275 |
| hardware_skeleton_r_elbow_confidence | 0 | 102 | 0 |
| hardware_skeleton_r_wrist_x | 412 | 0 | 2531 |
| hardware_skeleton_r_wrist_y | 412 | 0 | 2531 |
| hardware_skeleton_r_wrist_confidence | 0 | 412 | 0 |
| hardware_skeleton_l_shoudler_x | 238 | 0 | 2019 |
| hardware_skeleton_l_shoudler_y | 238 | 0 | 2019 |
| hardware_skeleton_l_shoudler_confidence | 0 | 238 | 0 |
| hardware_skeleton_l_elbow_x | 816 | 0 | 2269 |
| hardware_skeleton_l_elbow_y | 816 | 0 | 2269 |

Figure 19. Statistical Values for Some Skeletal Data Based on Three Metrics

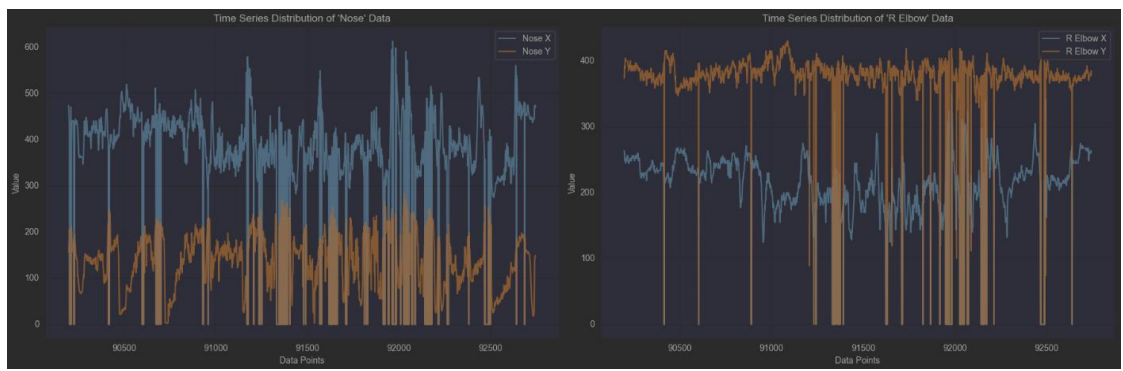

Figure 20. Filtered Results

Before addressing the outliers, a cursory glance at the time series variation might assist us in identifying the best approach.

Figure 21. Time Series Plot for Nose and Right Elbow

The time series plot reveals a rather bleak situation with the data. Even though the nose and right elbow data are relatively complete and reliable within the entire motion data set, addressing this scenario is challenging. We don't have sufficient information to recover the authentic data. Simply resorting to interpolation or data smoothing isn't a viable option, as it might exacerbate the data situation. Consequently, we opted to employ the ARIMA model from time series analysis to handle this data.

The ARIMA (Autoregressive Integrated Moving Average) model is a statistical approach specifically designed for time series data analysis and forecasting. This model combines autoregressive (AR), integrated (I), and moving average (MA) methods to capture time dependencies and structured patterns in the data. Figure 22 visualizes our post-processing results.

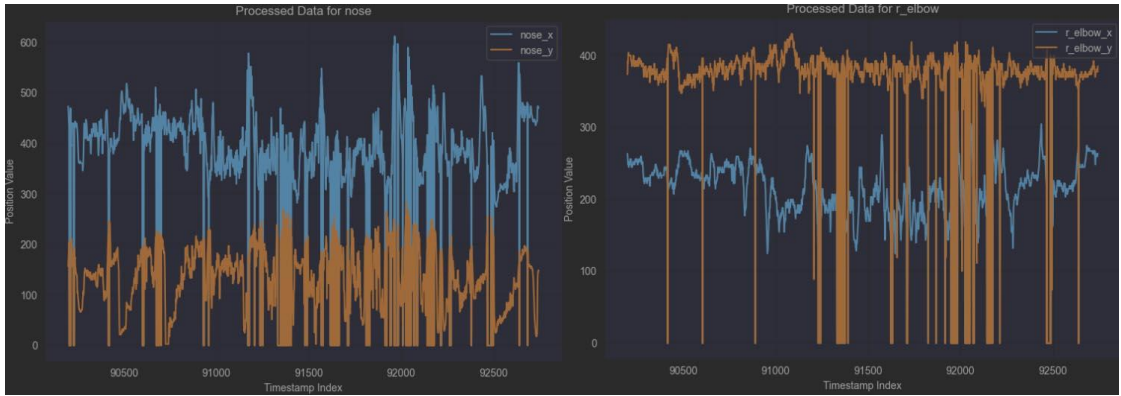

Figure 22. Time Series Plot for Two Skeletal Data Points Post ARIMA Model Processing

The data situation hasn't shown significant improvement. We attempted adjusting the ARIMA model parameters. Here, p denotes the number of autoregressive terms, reflecting the influence of past values on the current value; d signifies the number of times the data requires differencing to ensure data stationarity; and q represents the number of moving average terms, describing the error or noise in the data. The combination used in the above figure is the default parameters (p,d,q)=(1,1,10). The results remained largely unchanged. Given the potential significance of data authenticity, we decided to retain the original data.

**Time Series Comparison**:

Comparing the time series plots, we didn't discern any apparent patterns corresponding to the flow attribute. Although the positional data of body parts and the EDA data seem to have their inherent patterns, there's no clear direct association between them.

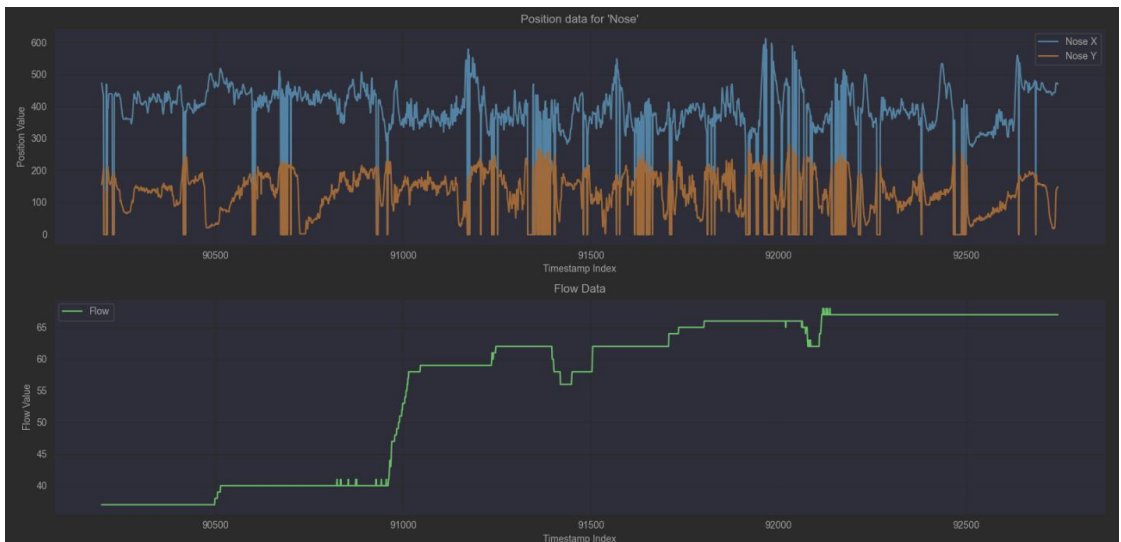

Figure 23. Time Series Comparison Plot for Nose's X & Y Coordinates and Flow

From the comparison, the musician's overall movement, as represented by the nose's position, doesn't show a clear correlation with their flow state. The dynamics of the nose's position might be influenced by various factors, and its interplay with the musician's psychological state (as represented by flow) isn't straightforward. The implications and potential reasons for these observations would warrant deeper exploration in further analysis.

## Experience 3. Cross-file Comparisons

In the two previous phases, we delved into the nuances of individual data sets. While the patterns and trends observed were not always explicit, it's possible to discern and compare patterns, trends, or differences across multiple files (or samples/musicians). Owing to potential inconsistencies and anomalies in other datasets, the EDA data, with its relative completeness and reliability, emerges as the ideal candidate for cross-file comparative analysis.

Initially, we embarked on a rudimentary exploration of the EDA data for different musicians by calculating descriptive statistics such as the mean, median, and standard deviation. The results underscored significant variations in EDA responses across musicians. Such disparities could be attributed to a musician's unique playing style, technique, or physiological attributes. A snippet of the statistical data for some musicians is depicted below, with a comprehensive version available in Appendix 3's code.

| session_id | count | mean | std | min | 25% | 50% | 75% | max |
|---|---|---|---|---|---|---|---|---|
| 4gSjKBs8A2Xxk52qk9aWDo | 3912.0 | 256.982106 | 12.227524 | 225.0 | 248.0 | 257.0 | 265.0 | 289.0 |
| 6qBvoFcD8f9Jp5FRHNdkm5 | 4005.0 | 230.951311 | 9.669869 | 215.0 | 223.0 | 230.0 | 237.0 | 259.0 |
| 9tVwbZHnSsLMASEnZa7xqe | 3933.0 | 228.577676 | 43.666018 | 155.0 | 190.0 | 221.0 | 271.0 | 305.0 |
| AeDezh7JLEukFn6ezmDAMt | 3922.0 | 223.278939 | 52.290828 | 108.0 | 216.0 | 231.0 | 249.0 | 319.0 |
| BPLGiqmkv36z2ymUTtwwfJ | 2552.0 | 509.255094 | 75.169493 | 207.0 | 507.0 | 526.0 | 543.0 | 654.0 |
| Bbwdm6nTkNGjQDrAze55Rs | 4073.0 | 232.038055 | 8.408778 | 214.0 | 226.0 | 231.0 | 236.0 | 255.0 |
| CQskrJ6vZAEeqyCRGfB4UV | 4056.0 | 633.189596 | 65.148729 | 474.0 | 591.0 | 620.0 | 668.0 | 880.0 |
| FuakZoP9kKigYHd5WKjVVX | 4080.0 | 359.267892 | 12.182141 | 324.0 | 352.0 | 359.0 | 367.0 | 400.0 |
| Jn3VvBWcnDESzN9gUTh3bN | 2548.0 | 467.442308 | 84.651605 | 318.0 | 377.0 | 490.0 | 538.0 | 649.0 |
| JqQsbJgaV8MdMKJRyUjuXR | 4126.0 | 353.905477 | 40.174767 | 309.0 | 327.0 | 341.0 | 361.0 | 483.0 |

Figure 24. EDA Statistical Data Across Musicians

To visually elucidate these variations, we employed box plots to render the EDA data for each musician. The box plots lucidly delineate the data distribution, medians, and potential outliers for every musician.

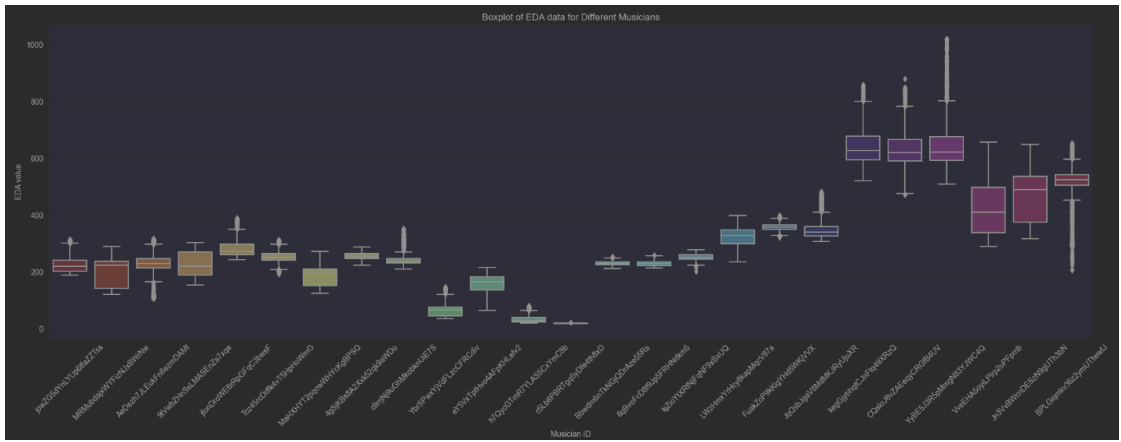

Figure 25. Box Plots of EDA Data Across Musicians

Distinct disparities are evident in the EDA data distributions among musicians. Each musician exhibits unique medians, quartiles, and potential outliers.

To reaffirm the pronounced disparities in patterns among musicians, we utilized Analysis of Variance (ANOVA) to statistically ascertain significant differences in EDA data across musicians. The results of the ANOVA are as follows:

F-statistic: 58725.45

p-value: 0.0

The obtained p-value, rounded off or considered as zero by computational software, insinuates that the inter-group variances are highly significant. This corroborates our initial observations: distinct musicians exhibit unique physiological responses, as manifested in their EDA data.

To more distinctly discern the differences in reactions between various musicians, we selected the top five musicians with the highest correlation. We then showcased the overlay of their EDA and 'flow' state time series within each chorus id segment. This approach allows for a clearer individualized perspective on the variations between different musicians.

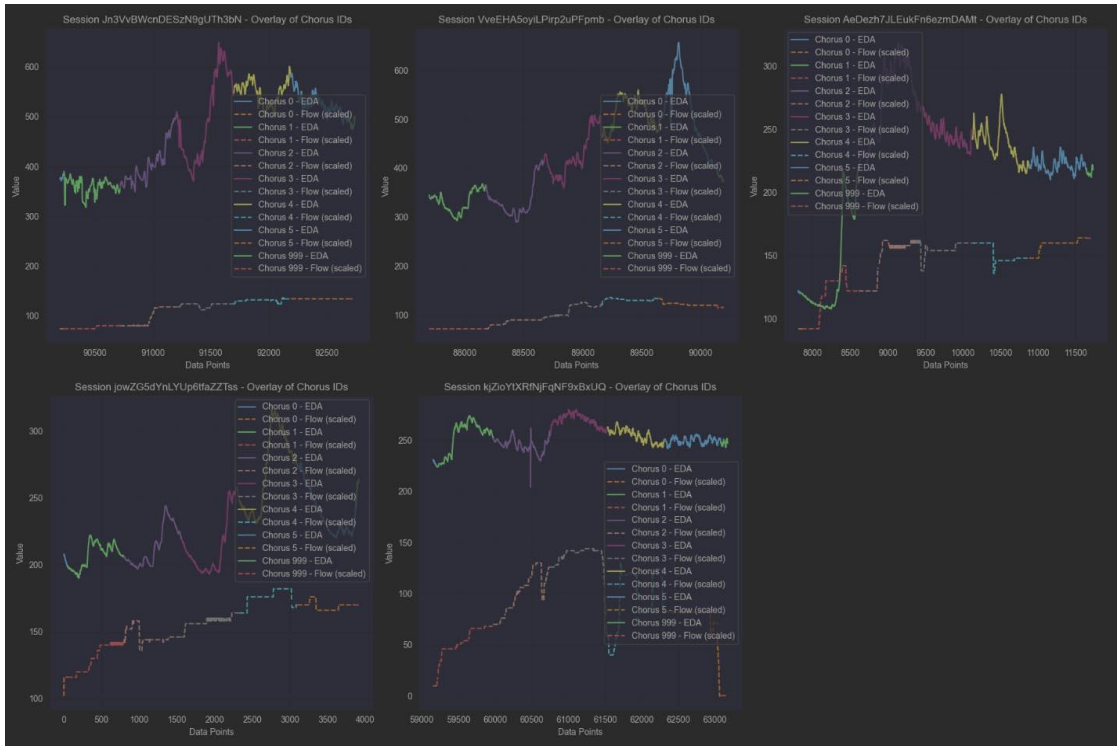

Figure 26. Time Series Overlay of EDA and 'flow' for the Top 5 Musicians with the Highest Correlation, Segmented by chorus id

From these graphs, we can distinctly observe the trends of EDA and 'flow' for different musicians across various choruses. Broadly speaking, in certain instances, we can discern a relationship between EDA and the 'flow' state; for example, when EDA rises, the 'flow' state also tends to increase. However, this isn't consistently the case, as one musician's EDA and 'flow' state appears to be inversely related. While the EDA remains relatively stable across different chorus id for certain musicians, for most, there are evident fluctuations. For the majority of the musicians, the 'flow' state varies across different chorus id, but no clear pattern emerges. This variability might be tied to their distinct reactions to the backing track.

Through the analysis delineated above, it's unequivocally evident that distinct musicians exhibit significant variations in their physiological responses, as captured by the EDA data. These disparities might stem from individual musicians' unique techniques, playing styles, experiences, or other intrinsic variances. This revelation holds paramount significance in further understanding the physiological and bodily responses of musicians during performances, especially in relation to their experiences of the "flow" state.

Moreover, this signifies that a singular, macroscopic analytical approach might not suffice to capture all pivotal patterns and trends during cross-file analyses. A deeper, more personalized analysis might be imperative for each musician to holistically comprehend their physiological responses during performances.

In summation, cross-file comparative analysis furnishes invaluable insights into the variances between distinct musicians. In subsequent research, we intend to delve deeper into individualized scenarios for a more comprehensive analysis.

# Experience 4. Focused Single-File Analysis

After the explorations and analyses in the first three stages, we decided to delve deeper into a more specific analysis of individual files, aiming to better understand the physiological reactions of musicians while performing specific tasks. We continue to focus on the musician selected in the second stage. Our main emphasis will be on the EDA data since, as established in the previous analysis, it is deemed to be a relatively reliable and comprehensive set of physiological data. With that in mind, we proceeded with further analysis.

## Audio Data

To achieve a more granular time segmentation, we analyzed the backing track audio file to extract node information such as beats or bars.

Here's the essential information we extracted regarding the backing track:

Estimated Tempo: The tempo, representing the speed or rhythm of music, is typically expressed in beats per minute (BPM). Here, the estimated tempo is 60.09 BPM, suggesting roughly one beat per second, indicative of a relatively slow pace, potentially characteristic of jazz music.
Sampling Rate: 22050 Hz. This implies that the audio file comprises 22,050 sample points every second, a common sampling rate, especially for certain audio processing tasks.
Duration: 331.5 seconds. This means the track has a total duration of approximately 5 minutes and 31.5 seconds.
Additionally, we obtained specific time nodes for beats and bars. This allows us to segment the music into bars and average or sum the EDA and Flow data within each bar. Detailed time nodes can be referred to in Appendix 4.

## Fine-grained Time Analysis

We overlaid the EDA time series with the flow time series for one musician across five choruses. In the flow attribute, the value of flow is higher for each chorus than the previous one. However, given the imprecise nature of self-reported flow values, fluctuations within a single chorus for flow aren't pronounced. For EDA, although not every chorus has a higher value than the previous one, overall, the mean EDA in the later choruses is higher than in the earlier ones. Regrettably, no distinct patterns are discernible from the current data.

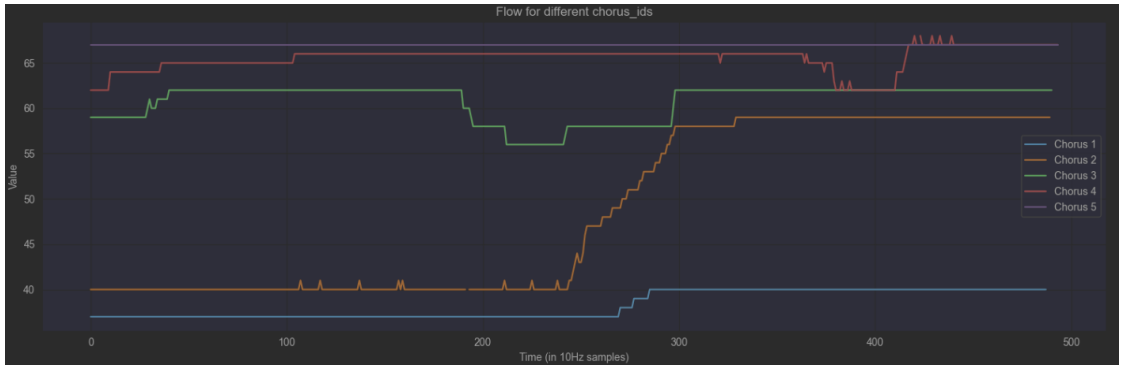

Figure 27. Flow Time Series for 5 choruses

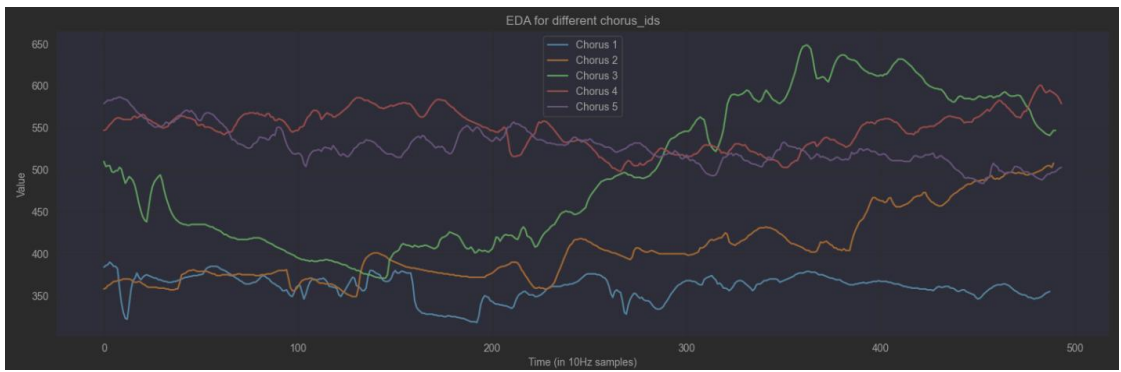

Figure 28. EDA Time Series for 5 choruses

## Clustering Analysis

Our objective was to explore the differences in physiological reactions (EDA data) of a specified musician across five different performances (i.e., five choruses). For this, we employed clustering analysis, an unsupervised machine learning technique that groups data points such that those within the same group are more similar to each other than to those in other groups. We used the KMeans clustering method and experimented with different cluster numbers. In the analysis, three clusters emerged as the optimal choice. Their respective sizes were 38, 30, and 13, leading us to hypothesize:

Cluster 0: Encompassing the majority of bars, this suggests these bars share statistical similarities in EDA. Given its prevalence, it might signify a "baseline" or "norm" – perhaps the musician's default state for most of the track. These bars might represent the musician's consistent style and emotional expression.

Cluster 1: Bars in this cluster have a different EDA reaction than those in Cluster 0. These bars, being moderately numbered, might represent sections where the musician maintains a consistent and steady style and emotion, without significant ups and downs.

Cluster 2: The smallest cluster, bars here might represent sections where the musician showcases specific emotions or technical prowess. These bars, being fewer, might represent special or challenging parts of the track.

Given the characteristic of the piece, comprising five consecutive choruses, the musician might have experimented with different playing styles or emotional expressions in each chorus. The initial chorus could be an "exploration" or "warm-up" phase, middle choruses might peak in intensity or technical showcases, and the final chorus might signify a gradual winding down or return to calm.

Basing on the clustering results, bars in Cluster 0 might predominantly belong to one or two choruses, representing sections where the musician showcases technical and emotional peaks. In contrast, bars in Cluster 2 might be spread across all choruses, acting as the "baseline" throughout.

Of course, these are conjectures based on the data. A genuine interpretation might require deeper musical theoretical knowledge and an understanding of the musician's style.

# Discussion

In this study, although we haven't delved into every detail, we've established a fundamental understanding and insight into the dataset. Our data preprocessing and preliminary analysis provide a framework for future researchers. While our analysis might not be the most exhaustive or precise, it offers a starting point for further research. Future researchers can build upon our work, combining more data and advanced analytical methods for a deeper exploration.

## Review and Evaluation of the Analysis Process

Throughout this research, we've adopted a series of data analysis steps to probe into the physiological reactions of musicians during performances. Spanning from macro to micro, our analytical journey covered multiple layers, enabling a comprehensive understanding of the musicians' physiological reactions. However, our analysis methods were relatively straightforward. Even though we tried to diversify our approach, many attempts turned out futile due to the exploratory nature of the experiment and the unique characteristics of each dataset. Additionally, constrained by time, our deep analysis was limited to one musician, which restricted the universality of our conclusions. Moreover, we only employed one machine learning method. While KMeans clustering was applied, it has its limitations, such as the challenge of intuitively selecting the right number of clusters and potential influence from initial conditions. Future analyses could benefit from incorporating data from more musicians to enhance universality and accuracy, and from leveraging more complex machine learning methods, such as hierarchical clustering or deep learning.

# Association of Music with Psychological State and Multimodal Physiological Data

Music has always been perceived as a potent emotional and cognitive stimulus, capable of invoking a plethora of physiological reactions. For instance, a rapid rhythm or tense chord might elevate an audience's physiological arousal—observable through increased EDA or heart rate variations. In contrast, a mellow melody or harmonious chord could induce a relaxed physiological state, potentially reflected in the electroencephalogram (EEG) alpha wave activity. Hence, music can be seen as a multifaceted stimulus, influencing audiences on multiple physiological levels. Although both music and flow are intertwined with physiological reactions, combining the two remains a relatively uncharted domain. In theory, when a musician enters a flow state, their physiological indicators, such as EDA or EEG, might manifest patterns consistent with states of relaxation or intense focus. However, given the intricate nature of musical creation—encompassing cognitive, emotional, and motor facets—predicting or interpreting these physiological shifts can be challenging. To better comprehend the nexus between music, flow, and physiological states, the future direction might lie in multimodal exploration, potentially merging conventional physiological measurement techniques, such as EDA and EEG, with advanced neuroimaging technologies like functional MRI. Moreover, considering music's diversity, researchers might need to study various music genres, cultures, and backgrounds to capture a broader spectrum of physiological and psychological responses. Lastly, technological advancements, like wearables or real-time physiological monitoring tools (Johnson and Picard, 2020), might pave the way for deeper exploration of musicians' flow experiences and physiological reactions in real-world scenarios.

## Individual Variability Among Musicians

The realm of musical performance mirrors the diversity inherent in human society. Even when subjected to similar training, no two musicians manifest identical physiological or psychological responses during their performances. This individual variability is influenced by an array of factors, ranging from genetic makeup, past experiences, and training methodologies to the pressures of everyday life. Such differences might manifest in various ways. For instance, genetic variations might lead to differences in muscle structures, neural pathways, or even how the body responds to stress. One musician, for example, might experience an accelerated heartbeat due to stage anxiety, while another might perspire more profusely. A musician trained in classical music might approach a piece differently from someone with a jazz background. Their technique, posture, and even emotional connection to the music might differ based on their training, leading to varied physiological responses. The mental state of musicians can significantly influence their physiological reactions. Personal stressors, mood, or even levels of motivation can result in distinct physiological patterns. As we continue to explore this database, future considerations could include segmental analyses, mixed-effects models, and the introduction of additional control variables. This also raises considerations for refining data collection methodologies in the database.

While individual differences introduce constraints and challenges, they also add tremendous value to our research. The diverse reactions and responses mean that our dataset is rich and varied, offering multifaceted insights for musicians when scaled up.

In conclusion, the individual differences among musicians, while complicating our analysis, also enrich our research findings. By acknowledging and accommodating these differences, we can paint a more comprehensive and authentic picture of the physiological and psychological landscape of musical performance.

# Database Limitations, Challenges, and Real-world Musical Applications

Throughout this research, we heavily relied on the database for abundant raw data. However, as we delved deeper, we encountered certain limitations and challenges of the database, which, to an extent, impeded the depth of our analysis and the reliability of our conclusions.

## Sampling Rate Issues

The database's sampling rate is pivotal for certain types of data analyses. For instance, standard EEG sampling rates usually range between 250-500 Hz or even higher(Weiergräber et al., 2016). Such rates can capture various brain frequency bands, such as $\alpha$, $\beta$, and $\gamma$ waves. However, the database's sampling rate was relatively low, which might lead to the omission of high-frequency brain activities. Due to the reduced sampling rate, signals with frequencies above 3.84 Hz (based on our single-file analysis of the musician "Jn3VvBWcnDESzN9gUTh3bN") could be misinterpreted as low-frequency signals, a consequence of the Nyquist theorem (Weiergräber et al., 2016). This reduced sampling rate might cause EEG signals to appear more synchronized, potentially overlooking rapid changes within short time frames. This limitation restricted our in-depth exploration of the relationship between music and brain activity, especially in the frequency domain.

## Data Confidence Issues

The database encompasses vast skeletal data, which, in theory, could provide invaluable insights into the musician's posture and movements. However, much of this data exhibited low confidence, indicating potential inaccuracies or noise. This complicates data preprocessing and might affect the accuracy of our analytical outcomes.

### Real-world Musical Applications

At the project's end, an intriguing discovery emerged. While the dataset's various drawbacks made pattern discovery challenging from a data science perspective, three musicians trained neural networks on this dataset, eventually feeding it to AI robots, which produced music every second. Relative to our research methods, neural networks and deep learning models have proven to excel at handling complex, non-linear, and high-dimensional data, perhaps extracting meaningful features and learning the intrinsic structure of music. It's also possible that musicians didn't wholly rely on data during model training; their musical knowledge, intuition, and experience might have guided them in model design, feature selection, and post-processing. Another significant factor might be that the beauty of music is subjective; occasionally, imperfect, raw, or rough elements might add unique charm and depth to a piece. Such discoveries suggest that even under imperfect data conditions, a fusion of innovation and technology might yield unexpected results. Thus, this dataset remains a precious resource, offering a unique opportunity to explore the relationship between music, physiology, and psychology. This situation serves as a potent reminder that in both research and art, one shouldn't easily abandon or doubt data. Instead, one should seize every opportunity, believing that the amalgamation of technology, knowledge, and creativity can lead to genuinely valuable outcomes.

# Conclusion

This study delves into the physiological reactions of musicians during performances, especially their responses in a flow state. By integrating an analysis of physiological data, specific musical performances, and the musicians' experiences, we've gleaned a series of intriguing findings and speculations.

Firstly, we've come to recognize that there might be some form of correlation between music and physiological activity. Although no explicit patterns were discerned, potential connections and trends were observable from the data. These initial discoveries pave the way for future research, hinting that a more detailed and comprehensive approach might be required to further probe these relationships. These preliminary findings also provide a potential scientific foundation for music therapy and other music-related applications, affirming that music can not only touch the soul but might also be related to physical reactions.

Secondly, the flow state has been verified as a unique psychological and physiological state, associated with heightened focus, a balance between skills and challenges, and a distortion in the perception of time. Our findings suggest that when musicians enter a flow state, their physiological indicators might manifest patterns consistent with this psychological state. This provides a fresh perspective for understanding and fostering the flow state, suggesting that physiological indicators could be monitored to identify and enhance flow experiences.

However, this research also unveiled several pivotal challenges. Due to the low sampling rate of the database and the confidence issues with skeletal data, we encountered difficulties analyzing EEG and other high-frequency data. This underscores the importance of selecting and employing high-quality data in future research. Moreover, given that flow is a complex psychological state, capturing it solely through a few physiological indicators might be insufficient. This calls for a more holistic approach in future research, incorporating a range of physiological, cognitive, and emotional measurement techniques.

In summary, this study offers preliminary insights into the interplay between music, flow, and physiology. Although we faced some challenges, we believe that with further research and technological advancements, we'll achieve a deeper understanding of the interactions between these phenomena, providing valuable knowledge for domains like music education, therapy, and human-machine interactions.

# Appendix

## Appendix 1

Please refer to the Appendix 1 code file in the link.
https://1drv.ms/f/s!AvU6PlzyQVvWliC19K2z7Hs-dLYN?e=iSRZcb

## Appendix 2

Please refer to the Appendix 2 code file in the link.
https://1drv.ms/f/s!AvU6PlzyQVvWliC19K2z7Hs-dLYN?e=iSRZcb

## Appendix 3

Please refer to the Appendix 3 code file in the link.
https://1drv.ms/f/s!AvU6PlzyQVvWliC19K2z7Hs-dLYN?e=iSRZcb

# Appendix 4

Please refer to the Appendix 3 code file in the link.
https://1drv.ms/f/s!AvU6PlzyQVvWliC19K2z7Hs-dLYN?e=iSRZcb

Beats Time: [  0.55727891  1.55573696  2.55419501  3.55265306  4.55111111
   5.54956916  6.54802721  7.50004535  8.54494331  9.54340136
  10.54185941 11.54031746 12.53877551 13.53723356 14.53569161
  15.53414966 16.55582766 17.55428571 18.55274376 19.55120181
  20.54965986 21.54811791 22.54657596 23.54503401 24.54349206
  25.54195011 26.54040816 27.53886621 28.53732426 29.53578231
  30.53424036 31.53269841 32.55437642 33.55283447 34.55129252
  35.54975057 36.54820862 37.54666667 38.54512472 39.54358277
  40.54204082 41.54049887 42.53895692 43.53741497 44.53587302
  45.55755102 46.55600907 47.53124717 48.55292517 49.55138322
  50.54984127 51.52507937 52.54675737 53.54521542 54.54367347
  55.54213152 56.54058957 57.53904762 58.53750567 59.53596372
  60.53442177 61.53287982 62.55455782 63.55301587 64.55147392
  65.54993197 66.54839002 67.54684807 68.54530612 69.54376417
  70.54222222 71.54068027 72.53913832 73.53759637 74.53605442
  75.55773243 76.55619048 77.53142857 78.55310658 79.55156463
  80.55002268 81.54848073 82.54693878 83.54539683 84.54385488
  85.54231293 86.54077098 87.51600907 88.53768707 89.53614512
  90.53460317 91.53306122 92.55473923 93.55319728 94.55165533
  95.55011338 96.54857143 97.54702948 98.54548753 99.54394558
 100.54240363 101.54086168 102.53931973 103.53777778 104.53623583
 105.53469388 106.55637188 107.55482993 108.55328798 109.52852608
 110.55020408 111.54866213 112.54712018 113.54557823 114.54403628
 115.51927438 116.54095238 117.53941043 118.53786848 119.53632653
 120.53478458 121.55646259 122.55492063 123.55337868 124.55183673
 125.55029478 126.54875283 127.54721088 128.54566893 129.54412698
 130.54258503 131.54104308 132.53950113 133.53795918 134.53641723
 135.48843537 136.53333333 137.53179138 138.55346939 139.52870748
 140.55038549 141.54884354 142.54730159 143.54575964 144.54421769
 145.54267574 146.54113379 147.53959184 148.53804989 149.53650794
 150.53496599 151.53342404 152.53188209 153.53034014 154.55201814
 155.52725624 156.54893424 157.54739229 158.54585034 159.54430839
 160.54276644 161.54122449 162.53968254 163.53814059 164.53659864
 165.53505669 166.55673469 167.53197279 168.53043084 169.52888889
 170.52734694 171.52580499 172.54748299 173.52272109 174.54439909
 175.51963719 176.54131519 177.53977324 178.53823129 179.53668934
 180.53514739 181.5568254  182.55528345 183.50730159 184.52897959
 185.52743764 186.54911565 187.5475737  188.54603175 189.5444898

```
190.54294785 191.5414059  192.53986395 193.538322   194.53678005
195.5352381  196.53369615 197.5321542  198.53061224 199.55229025
200.5507483  201.54920635 202.5476644  203.54612245 204.5445805
205.54303855 206.5414966  207.53995465 208.5384127  209.53687075
210.5353288  211.53378685 212.5322449  213.53070295 214.50594104
215.550839   216.54929705 217.5477551  218.5229932  219.47501134
220.54312925 221.5415873  222.54004535 223.5385034  224.56018141
225.3354195  226.53387755 227.5323356  228.53079365 229.55247166
230.55092971 231.54938776 232.5478458  233.5230839  234.5447619
235.54321995 236.51845805 237.54013605 238.46893424 239.5138322
240.5355102  241.53396825 242.5324263  243.48444444 244.5293424
245.55102041 246.54947846 247.54793651 248.54639456 249.54485261
250.5200907  251.54176871 252.54022676 253.53868481 254.53714286
255.53560091 256.55727891 257.55573696 258.55419501 259.55265306
260.55111111 261.54956916 262.54802721 263.54648526 264.54494331
265.52018141 266.54185941 267.54031746 268.53877551 269.53723356
270.53569161 271.53414966 272.55582766 273.55428571 274.55274376
275.55120181 276.54965986 277.52489796 278.54657596 279.54503401
280.54349206 281.54195011 282.54040816 283.53886621 284.53732426
285.53578231 286.55746032 287.55591837 288.55437642 289.55283447
290.55129252 291.52653061 292.54820862 293.54666667 294.54512472
295.52036281 296.54204082 297.51727891 298.53895692 299.53741497
300.53587302 301.53433107 302.55600907 303.55446712 304.55292517
305.55138322 306.54984127 307.54829932 308.54675737 309.54521542
310.54367347 311.54213152 312.54058957 313.53904762 314.53750567
315.53596372 316.55764172 317.55609977 318.55455782 319.55301587
320.55147392 321.54993197 322.54839002]
Bars Time: [  0.55727891   4.55111111   8.54494331  12.53877551  16.55582766
  20.54965986  24.54349206  28.53732426  32.55437642  36.54820862
  40.54204082  44.53587302  48.55292517  52.54675737  56.54058957
  60.53442177  64.55147392  68.54530612  72.53913832  76.55619048
  80.55002268  84.54385488  88.53768707  92.55473923  96.54857143
 100.54240363 104.53623583 108.55328798 112.54712018 116.54095238
 120.53478458 124.55183673 128.54566893 132.53950113 136.53333333
 140.55038549 144.54421769 148.53804989 152.53188209 156.54893424
 160.54276644 164.53659864 168.53043084 172.54748299 176.54131519
 180.53514739 184.52897959 188.54603175 192.53986395 196.53369615
 200.5507483  204.5445805  208.5384127  212.5322449  216.54929705
 220.54312925 224.56018141 228.53079365 232.5478458  236.51845805
 240.5355102  244.5293424  248.54639456 252.54022676 256.55727891
 260.55111111 264.54494331 268.53877551 272.55582766 276.54965986
 280.54349206 284.53732426 288.55437642 292.54820862 296.54204082
 300.53587302 304.55292517 308.54675737 312.54058957 316.55764172
 320.55147392]
```